\documentclass[pra,twocolumn,superscriptaddress,showpacs,amsmath,amssymb]{revtex4-1}

\usepackage{graphicx}
\usepackage{mathrsfs}
\usepackage{bm}
\usepackage{amssymb}
\usepackage{amsmath}
\usepackage{dcolumn}
\usepackage{color}

\newcommand{\ket}[1]{| #1 \rangle}
\newcommand{\bra}[1]{\langle #1 |}
\newcommand{\ketbra}[2]{| #1 \rangle \langle #2 |}
\newcommand{\expect}[1]{\langle #1 \rangle} 
\def\H{{\rm H}}
\def\V{{\rm V}}

\def\00{\H\V}
\def\11{\V\H}

\begin{document}

\title{Heralded amplification of nonlocality via entanglement swapping for long-distance device-independent quantum key distribution}

\author{Yoshiaki~Tsujimoto}
\affiliation{Advanced ICT Research Institute, National Institute of Information and Communications
Technology (NICT), Koganei, Tokyo 184-8795, Japan\\}

\author{Chenglong~You}
\affiliation{Advanced ICT Research Institute, National Institute of Information and Communications
Technology (NICT), Koganei, Tokyo 184-8795, Japan\\}
\affiliation{Hearne Institute for Theoretical Physics and Department of Physics $\&$ Astronomy,
Louisiana State University, Baton Rouge, LA 70803, United States\\}

\author{Kentaro~Wakui}
\affiliation{Advanced ICT Research Institute, National Institute of Information and Communications
Technology (NICT), Koganei, Tokyo 184-8795, Japan\\}

\author{Mikio~Fujiwara}
\affiliation{Advanced ICT Research Institute, National Institute of Information and Communications
Technology (NICT), Koganei, Tokyo 184-8795, Japan\\}

\author{Kazuhiro~Hayasaka}
\affiliation{Advanced ICT Research Institute, National Institute of Information and Communications
Technology (NICT), Koganei, Tokyo 184-8795, Japan\\}

\author{Shigehito~Miki}
\affiliation{Advanced ICT Research Institute, National Institute of Information and Communications Technology~(NICT), Kobe, Hyogo 651-2492, Japan\\}
\affiliation{Graduate School of Engineering Faculty of Engineering, Kobe University, Kobe, Hyogo 657-0013, Japan\\}

\author{Hirotaka~Terai}
\affiliation{Advanced ICT Research Institute, National Institute of Information and Communications Technology~(NICT), Kobe, Hyogo 651-2492, Japan\\}

\author{Masahide~Sasaki}
\affiliation{Advanced ICT Research Institute, National Institute of Information and Communications
Technology (NICT), Koganei, Tokyo 184-8795, Japan\\}

\author{Jonathan~P.~Dowling}
\affiliation{Advanced ICT Research Institute, National Institute of Information and Communications
Technology (NICT), Koganei, Tokyo 184-8795, Japan\\}
\affiliation{Hearne Institute for Theoretical Physics and Department of Physics $\&$ Astronomy,
Louisiana State University, Baton Rouge, LA 70803, United States\\}

\author{Masahiro~Takeoka}
\affiliation{Advanced ICT Research Institute, National Institute of Information and Communications
Technology (NICT), Koganei, Tokyo 184-8795, Japan\\}




\begin{abstract}
To realize the practical implementation of device-independent quantum key distribution~(DIQKD), the main difficulty is that its security relies on the detection-loophole-free violation of the Clauser-Horne-Shimony-Holt~(CHSH) inequality, i.e. the CHSH value $S>2$, which is easily destroyed by the loss in transmission channels. 
One of the simplest methods to circumvent it is to utilize the entanglement swapping relay~(ESR). 
Here, we propose and experimentally test 
an improved version of the heralded nonlocality amplifier protocol based on the ESR, and numerically show that our scheme is much more robust against the transmission loss than the previously developed protocol. 
In the experiment, we observe that the obtained probability distribution is in excellent agreement with those expected by the numerical simulation with experimental parameters which are precisely characterized in a separate measurement. 
Moreover, we experimentally estimate the nonlocality of the heralded state after the transmission of 10~dB loss just before detection. 
It is estimated to be $S=2.104>2$, which indicates that our final state possesses strong nonlocality even with various experimental imperfections. Our result clarifies an important benchmark of the ESR protocol, and  paves the way towards the long-distance realization of the loophole-free CHSH-violation as well as DIQKD.
\end{abstract}

\maketitle

\section{Introduction}
\label{secI}
Nonlocality is one of the most interesting features of quantum mechanics which can be tested by the celebrated Bell inequality~\cite{bell1964js,RevModPhys.86.419}. 
Recently, it has been also pointed out that the system violating the Bell inequality in a loophole-free manner is directly related to quantum information applications such as device-independent quantum key distribution~(DIQKD)~\cite{PhysRevLett.98.230501,pironio2009device}.
Remarkably, DIQKD relaxes the requirements for the security proof, and 
allows the two users, Alice and Bob, to share an information-theoretic-secure secret key without making assumptions about internal workings of the physical devices. 
However, its practical implementation is challenging, 
since the security is solely based on the loophole-free violation of the Bell's inequality. 
One of the most formidable loopholes is the detection loophole~\cite{giustina2013bell,PhysRevLett.111.130406,PhysRevLett.115.250401,PhysRevLett.115.250402,PhysRevLett.120.010503,PhysRevLett.121.150402}, which necessitates the receiver to detect at least 2/3 of emitted photons~\cite{PhysRevA.47.R747}. 
That is, if a standard optical fiber at telecommunication wavelength 
with 0.2~dB/km-loss is used as a transmission channel, the achievable distance 
becomes less than 10~km even if photon detectors with unity detection efficiencies are employed. 

Several protocols to circumvent this difficulty have been proposed, such as the linear-optics-based heralded qubit amplifier~\cite{PhysRevLett.105.070501,PhysRevA.84.022325,PhysRevA.88.012327} and the heralding protocol with a  nonlinear process~\cite{PhysRevLett.106.120403}. 
When a single photon state~(such as a part of the entangled photon pair) is sent into a lossy channel, the state turns out to be a mixture of a single-photon state and a vacuum. 
These protocols can increase the fraction of qubit~(single photon) and suppress the vacuum fraction with a certain probability. 
Thus by applying them to the Bell state transmission, one can recover the lost Bell state with some success probability. 
Although a proof-of-principle of the heralded qubit amplification was experimentally demonstrated by S. Kocsis {\it et al.} in 2013~\cite{kocsis2013heralded}, 
the generated state after heralding still contains significant amount of vacuum such that the state loses its nonlocality. 

An alternative option is the linear-optical entanglement swapping relay~(ESR), which is widely used as 
an entangling operation of independently prepared photon pairs in the postselection manner~\cite{PhysRevLett.80.3891,halder2007entangling,jin2015highly,tsujimoto2018high}. 
While the ESR is simpler than the other schemes, at first, this method was not believed to work in the experiment {\it without} postselection 
\cite{PhysRevLett.105.070501,PhysRevLett.106.120403} such as DIQKD.
This is because if one applies the ESR to the entangled photons generated by the spontaneous parametric-down conversion~(SPDC), 
which is currently the most practical source of a photonic entangled state, even if the swapping is successful, the generated state~(without postlsection) is far from 
the two-qubit maximally entangled state~(less than 0.5 fidelity between them).  
Surprisingly, however, 
M.~Curty and T.~Moroder~\cite{PhysRevA.84.010304} showed that the ESR without postselection is, in fact, able to violate
the  Clauser-Horne-Shimony-Holt~(CHSH) inequality~\cite{PhysRevLett.23.880}, 
i.e. the CHSH value $S>2$. 
This was confirmed by the following numerical analysis  by K.~P.~Seshadreesan {\it et al.}~\cite{PhysRevA.93.042328}, 
which contains various practical imperfections and the multi-pair generation of the SPDC sources. 
These theoretical predictions show that even if the ESR state is not close to the ideal Bell state, it still shows nonlocality, which is useful for quantum protocols such as DIQKD. 
Related to this, not the CHSH inequality violation but the event-ready quantum steering was recently demonstrated 
by using the ESR~\cite{Westone1701230}.

In this paper, we show the following major progress in this direction: we propose an improved scheme of a heralded nonlocality amplifier based on the ESR, perform it experimentally, and estimate the nonlocality of the experimentally generated state by the ESR-based heralding, which shows the violation of the CHSH inequality even after transmitting through a channel with loss corresponding to the 50~km-optical fiber. 
More specifically, first we propose a modified scheme of the ESR heralding from the previous one~\cite{PhysRevA.84.010304,PhysRevA.93.042328}, and numerically compare their performances in detail. 
We show that when the total loss corresponds to the attenuation of the 100~km-optical fiber, 
the DIQKD key rate of our new scheme is almost 100 times larger than that of the previous one. 
Second, we perform a proof-of-principle experiment of the proposed scheme. 
Entangled photons from the SPDC sources are transmitted through lossy channels corresponding to the 50km-optical fiber 
and then the ESR heralding is performed. 
Although the detection efficiencies of our system are not in the range of directly observing
the violation of the CHSH inequality of the heralded state, the probability distributions 
obtained by the experiment are in excellent agreement with those independently obtained by our numerical model 
that includes imperfections. 
This feature allows us to estimate the nonlocality and the density matrix of the experimentally heralded state before the final detection. 
The estimated CHSH value is $S= 2.104$, which shows that the experimentally heralded state has significant nonlocality, while the fidelity of the heralded state to the two-qubit maximally entangled state is estimated to be 0.47.
This result indicates that we successfully amplified the nonlocality of the SPDC-based entangled photons, which are degraded by losses in the transmission channels, via the ESR. 
As far as we know, this is the first experiment recovering the nonlocality of the SPDC-based entangled photons after significant transmission losses. In light of the practicality of the SPDC-based entangled photons, 
our work paves a way to realize long-distance DIQKD by combining it with the state-of-the-art highly efficient photon detectors. 

The paper is organized as follows. 
In Sec.~\ref{sec1.5}, we briefly review the ESR-based heralded nonlocality amplifier in \cite{PhysRevA.84.010304} and then propose a modified scheme.  In Sec.~\ref{secII}, we describe our theoretical model and show the numerical result comparing these two schemes. 
The experimental setup and results are described in Sec.~\ref{secIII}.
In Sec.~\ref{secIV}, we discuss the density operator of the heralded state 
and Sec.~\ref{secV} concludes the paper.

\begin{figure}
 \begin{center}
 \includegraphics[width=\columnwidth]{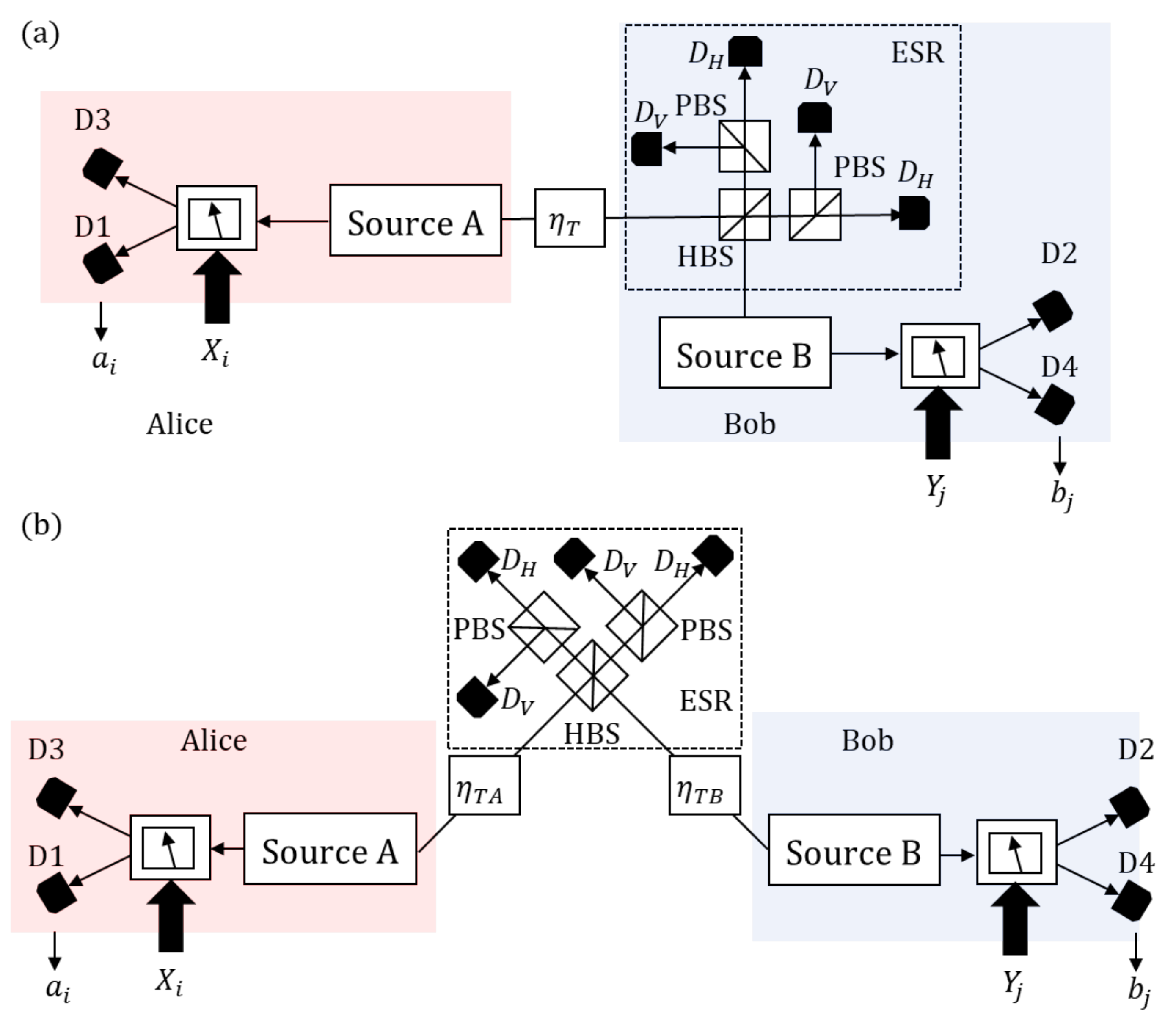}
  \caption{The schematic diagram of the ESR-based Bell-test experiment. 
	Linear optical Bell-state measurement~(BSM) is realized by 
	a half beamsplitter~(HBS) followed by polarization measurements using 
	two polarization beamsplitters~(PBSs).  
	Under the condition of the successful BSM at the ESR node, 
	Alice and Bob perform the polarization-measurement based on their measurement settings $X_i\in\{X_1, X_2\}$ and $Y_j\in\{Y_1, Y_2\}$, respectively. 
	By repeating the measurement, they calculate the CHSH value $S$.
  (a)~The ordinary configuration of the ESR-based DIQKD~(the SH scheme).
	(b)~The CH scheme we introduce.  
	The BSM is performed in the middle of Alice and Bob.
     \label{fig:ESR}}
 \end{center}
\end{figure}

\section{HERALDED NONLOCALITY AMPLIFICATION BY ENTANGLEMENT SWAPPING}
\label{sec1.5}

\begin{figure*}[t]
 \begin{center}
\scalebox{0.4}{\includegraphics{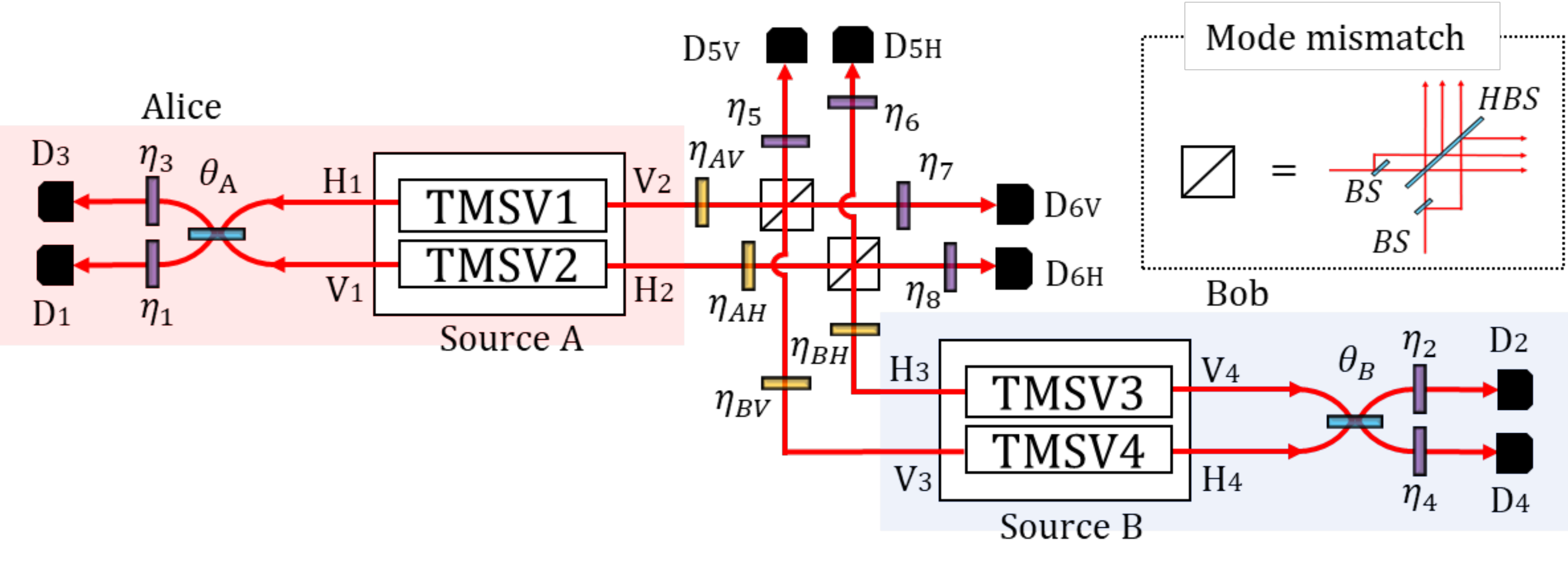}}
  \caption{The realistic model of the ESR-based Bell-test experiment. A pair of two-mode squeezed vacua~(TMSV)
	is used to prepare polarization entangled photon pairs. The linear optical Bell-state measurement is composed of 
	a half beamsplitter~(HBS) followed by polarization measurement at each output port. Alice~(Bob) set the angle of the polarizer to be $\theta_A~(\theta_B)$ and perform polarization measurements. 
	All of the photon detectors are the on-off type, single photon detectors with dark counts. 
     \label{fig:ESRREAL}}
 \end{center}
\end{figure*}

In this section, we review the ESR-based heralding scheme and then propose its improved version. 
The ESR-based heralding scheme proposed in Ref.~\cite{PhysRevA.84.010304} is illustrated in Fig.~\ref{fig:ESR}(a). 
Entangled photons are prepared at Alice's side by Source A. One of them is sent to Bob via a lossy optical channel with transmittance $\eta_T$, which easily destroys the nonlocality of the state. 
Bob prepares another entanglement source~(Source B) and perform the ESR by the Bell state measurement (BSM) to recover the lost nonlocality of the shared state between Alice and Bob. 
Since the ESR succeeds only probabilistically, this is a probabilistic protocol and we use the state only when heralded by the successful events of the ESR. 

In practice, the entangled sources A and B are based on the SPDC, which generates entangled photon pairs only probabilistically, and moreover, sometimes generates multiple pairs simultaneously. 
Then, by successful swapping, the heralded state $\hat{\rho}_{AB}$ mainly consists of the superposition of the following three events: 
(i)~two photon pairs from the source~A and no photon pair from source~B, (ii)~two photon pairs from the source~B and no photon pair 
from source~A, and 
(iii)~one photon pair from each of the sources~A and B. 
Clearly, (iii) is the desirable event, but 
the probability that the 
unwanted events (i) or (ii) occur is almost the same probability as (iii). 
Therefore, the fidelity of the heralded state~$\hat{\rho}_{AB}$ to the two-qubit maximally entangled states never exceeds 0.5, which was thought to be a reason that the generated state lost its nonlocality~\cite{PhysRevLett.105.070501,PhysRevLett.106.120403}. 
However, as shown by M.~Curty and T.~Moroder~\cite{PhysRevA.84.010304}, this $\hat{\rho}_{AB}$ still violates the CHSH inequality. 
That is, $\hat{\rho}_{AB}$ contains some nonlocality, although it is far from an ideal Bell state.

Here, we propose to modify the above scheme, which is illustrated in Fig.~\ref{fig:ESR}(b).  
The difference of it from the one in Fig.~\ref{fig:ESR}(a) is that the BSM is located not in Bob's side but in the middle of the channel 
and thus the channel is split into two with $\eta_{TA}$ and $\eta_{TB}$, respectively. 
This is possible since the BSM is not necessarily located inside Alice and Bob's systems. 
A similar configuration with single-photon sources was reported in Ref.~\cite{1803.07089}. 
Hereafter, we call this configuration the center-heralding (CH) scheme and the other one as the side-heralding (SH) scheme. 
These two are compared in detail in the next section, and as shown in there, the CH scheme shows much better performance than that of the SH scheme, especially when we take into account practical imperfections.  

\section{THEORETICAL ANALYSIS}
\label{secII}
\subsection{Model}
We first explain the procedure to generate a raw key using the ESR protocols in Fig.~\ref{fig:ESR}. 
Alice~(Bob) generates entangled photon pairs at the source~A(B). 
The linear optical BSM is performed at the ESR node, which 
is placed in Bob's system for the SH scheme and in the middle of Alice and Bob for the CH scheme. 
Under the condition that the BSM succeeds, 
Alice and Bob perform the polarization-measurements based on 
the measurement settings $X_i\in\{X_1, X_2\}$ 
and $Y_j\in\{Y_1, Y_2\}$, respectively. 
The measurement outcomes are binary, i.e., $a_i, b_j\in\{-1, +1\}$. By repeating the measurement, 
they calculate 
\begin{equation}
S=\expect{a_1b_1}+\expect{a_2b_1}+\expect{a_1b_2}-\expect{a_2b_2},
\label{eq:CHSH}
\end{equation}
where $\expect{a_ib_j}=P(a=b|X_i,Y_j)-P(a\neq b|X_i,Y_j)$. 
While the maximal value of $|S|$ 
is upper-bounded by 2 in the framework of a local realism theory, 
quantum mechanics allows $|S|$ to take the maximal value of $2\sqrt{2}$, 
which is known as the Tsirelson bound~\cite{cirel1980bs}. 
When Alice and Bob perform DIQKD, Alice chooses another measurement basis $X_0$, and 
the raw key is generated by the outcomes under the measurement setting of $\{X_0,Y_1\}$. 
The lower bound of the key rate $K$ is represented by Devetak-Winter formula~\cite{devetak2005distillation} as 
\begin{equation}
K\geq1-h(Q)-\chi(S), 
\label{eq:keyrate}
\end{equation}
where $Q$ is qubit error rate~(QBER) which is defined by $P(a\neq b|X_0,Y_1)$, and 
\begin{equation}
\chi(S)=h\left[\frac{1+\sqrt{(S/2)^2-1}}{2}\right].  
\end{equation}
Here, $h(\cdot)$ is the binary entropy defined by $h(x)=-x\mathrm{log}_2x-(1-x)\mathrm{log}_2(1-x)$.

As a realistic model of the ESR-based Bell-test with SPDC sources, we construct a theoretical model similar to the one 
introduced in Ref.~\cite{PhysRevA.93.042328}, as shown in Fig.~\ref{fig:ESRREAL}. 
Each entangled photon pair source consists of a pair of two-mode squeezed vacuum~(TMSV). 
The Hamiltonian is given by $\hat{H}=i\hbar(\zeta_{1(4)}\hat{a}^\dagger_{H_{1(3)}}\hat{a}^\dagger_{V_{2(4)}}+\zeta_{2(3)}\hat{a}^\dagger_{V_{1(3)}}\hat{a}^\dagger_{_{H2(4)}})+\rm{h.c.}$ for 
source~A~(B), 
where $\hat{a}_{ij}^\dagger$ is the photon-creation operator of the $i$-polarized single photon in mode $j$ 
which satisfies the commutation relation  $[\hat{a}_{ij},\hat{a}_{kl}^\dagger]=\delta_{ik}\delta_{jl}$. 
$H$ and $V$ denote the horizontal and vertical polarizations,
respectively.
$\zeta_k=|\zeta_k|e^{i\phi_k}$ is the coupling constant 
of TMSV$k$ ($k\in\{1,2,3,4\}$), which is proportional to the complex amplitude of each pump. 
In the following, $\phi_k$ is fixed as $\phi_1=\phi_2=\phi_3=0$ and $\phi_4=\pi$, which means that, 
when $|\zeta_1|^2=|\zeta_2|^2$ and $|\zeta_3|^2=|\zeta_4|^2$, 
the two-qubit components of the generated state form $\ket{\Psi^+}_{12}:=(\ket{HV}_{12}+\ket{VH}_{12})/\sqrt{2}$ for source~A, and 
$\ket{\Psi^-}_{34}:=(\ket{HV}_{34}-\ket{VH}_{34})/\sqrt{2}$ for source~B. 
Here, $\ket{H}_j:=\hat{a}^\dagger_{H_j}\ket{0}$ and $\ket{V}_j:=\hat{a}^\dagger_{V_j}\ket{0}$ denote the $H$- and $V$- polarization states of a single photon in mode $j$, respectively. 
At the ESR node, we perform the partial Bell-sate measurement using linear optics. 
We adopt the projection onto $\ket{\Psi^-}$, which is realized by detecting the two-fold coincidence between ($\mathrm{D_{5H}}\cap \mathrm{D_{6V}}$) or ($\mathrm{D_{5V}}\cap \mathrm{\mathrm{D_{6H}}}$). 
The successful operation of the ESR in the two-qubit system is described by 
\begin{eqnarray}
_{23}\bra{\Psi^-}\ket{\Psi^+}_{12}\ket{\Psi^-}_{34}&=&_{23}\bra{\Phi^+}\hat{Z}_2\hat{X}_2\hat{X}_2\hat{X}_3\hat{Z}_3\ket{\Phi^+}_{12}\ket{\Phi^+}_{34}\nonumber  \\
&=&_{23}\bra{\Phi^+}\hat{Z}_3\hat{X}_3\hat{Z}_3\ket{\Phi^+}_{12}\ket{\Phi^+}_{34}\nonumber  \\
&=&-_{23}\bra{\Phi^+}\hat{X}_3\ket{\Phi^+}_{12}\ket{\Phi^+}_{34}\nonumber  \\
&=&-\frac{1}{2\sqrt{2}}\hat{X}_4\sum_{j,k\in\{H,V\}}\delta_{j,k}\ket{j}_{1}\ket{k}_{4}\nonumber\\
&=&-\frac{1}{2}\ket{\Psi^+}_{14}, 
\end{eqnarray}
where $\hat{Z}:=\ketbra{H}{H}-\ketbra{V}{V}$ and $\hat{X}:=\ketbra{H}{V}+\ketbra{V}{H}$ 
are Pauli operators, and $\ket{\Phi^+}:=(\ket{HH}+\ket{VV}/\sqrt{2}$. 
The polarizer with angle $\theta$ works as a polarization-domain beamsplitter mixing the $H$ and $V$ modes whose 
transmittance and reflectance are $\mathrm{cos}^2\theta$ and $\mathrm{sin}^2\theta$, respectively. 
Under the condition of the two-fold coincidence between ($\mathrm{D_{5H}}\cap \mathrm{D_{6V}}$) or ($\mathrm{D_{5V}}\cap \mathrm{\mathrm{D_{6H}}}$), 
Alice~(Bob) chooses her~(his) angle from $\theta_A=\{\theta_{A0}, \theta_{A1},\theta_{A2}\}~(\theta_B=\{\theta_{B1},\theta_{B2}\})$, respectively, and performs polarization measurements.
The losses in the transmission channels are represented by 
$\eta_{AH}$, $\eta_{AV}$, $\eta_{BH}$, and $\eta_{BV}$.~(Thus, the SH scheme can be simulated by setting $\eta_{BH}=\eta_{BV}=0$.)
The local system losses including the imperfect quantum efficiencies of the detectors 
are modeled by inserting virtual loss materials denoted by $\eta_l$ for $l\in\{1,\cdots,8\}$.
We consider that all of the detectors are on-off type single-photon detectors, 
which only distinguish between vacuum~(off: no-click) and non-vacuum~(on: click).
the dark-count probability $\nu$, which is a false click of the detector, is also taken into account in the model.
The mode-mismatch between Alice's TMSV and Bob's TMSV is modeled 
by inserting virtual beamsplitter~(BS) whose transmittance is $T_\mathrm{mode}$ in each input port of the half beamsplitter~(HBS) at the ESR node 
as shown in the inset of Fig.~\ref{fig:ESRREAL}.
In other words, two virtual BSs divide the mode of the each TMSV into two parts: the mode which interferes with probability amplitude $\sqrt{T_\mathrm{mode}}$
and that does not with probability amplitude $\sqrt{1-T_\mathrm{mode}}$. 
The experimental value of $T_\mathrm{mode}$ can be determined by performing the Hong-Ou-Mandel interference experiment~\cite{PhysRevLett.59.2044,Tsujimoto:15,PhysRevA.87.063801}.

 \begin{figure}[t]
 \begin{center}
 \includegraphics[width=\columnwidth]{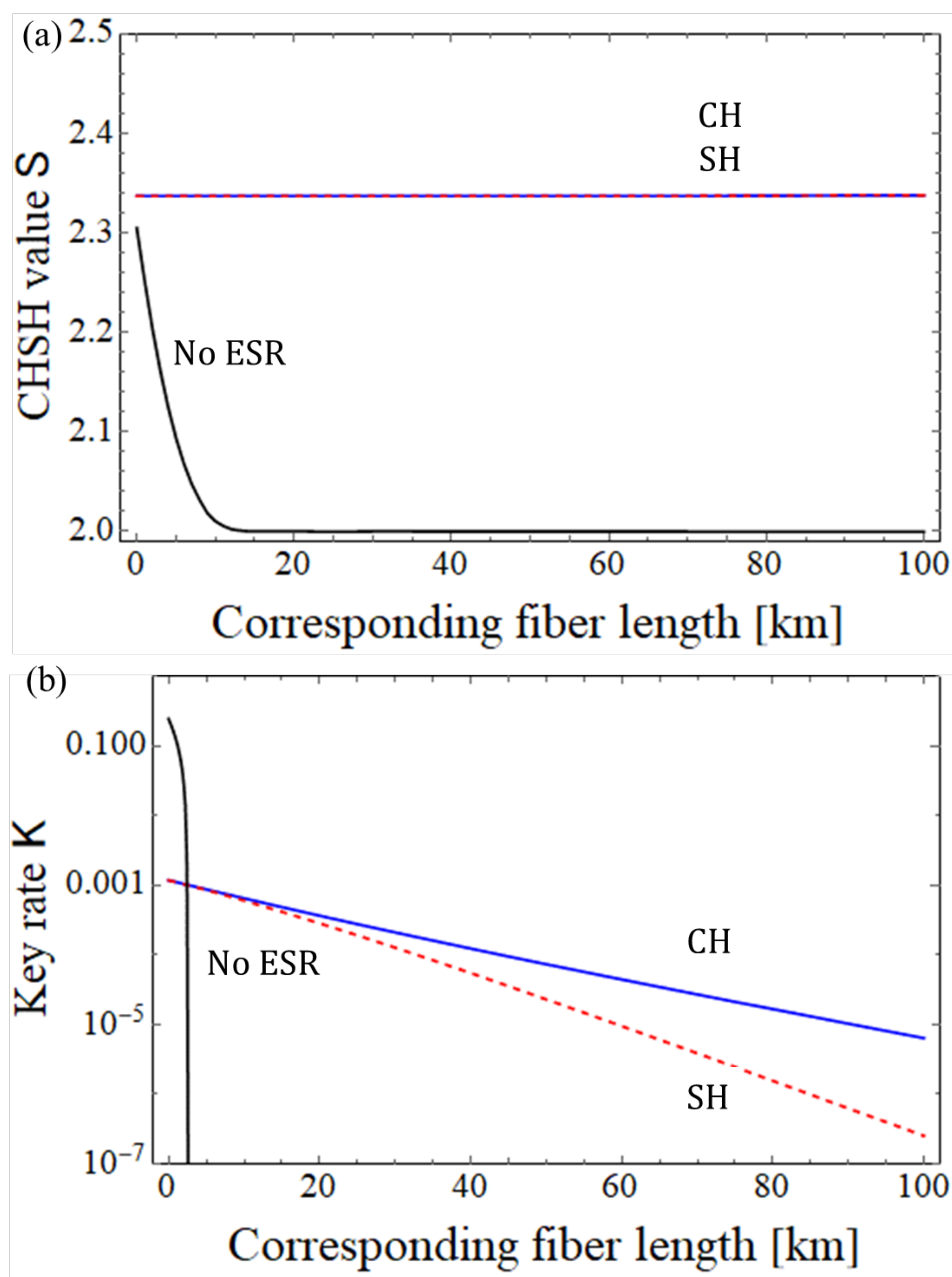}
  \caption{(a), (b)~The corresponding fiber length $L$ vs $S(K)$ in the ideal situation~($\forall l$ $\eta_l$=1, $T_\mathrm{mode}=1$, and $\nu=0$). 
	The blue solid curve and the red dashed curve are $S(K)$ for the CH and for the SH scheme, respectively. 
	For each of (a) and (b), the black solid curve corresponds the case where the ESR node is absent. }
\label{fig:d=0}
 \end{center}
\end{figure}

 \begin{figure}[t]
 \begin{center}
 \includegraphics[width=\columnwidth]{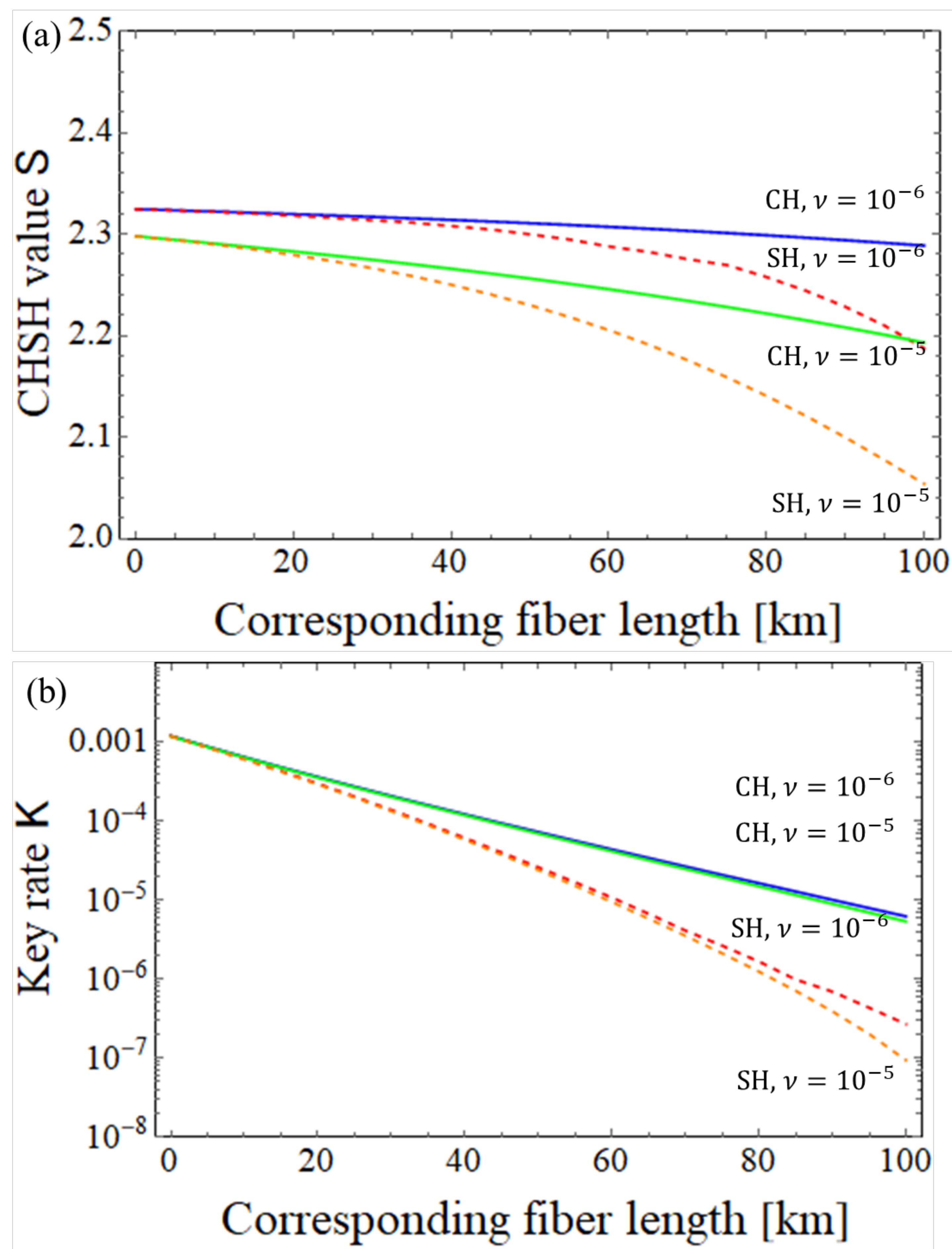}
  \caption{(a), (b)~$L$ vs $S(K)$ with dark counts~($\forall l$ $\eta_l$=1, $T_\mathrm{mode}=1$, and $\nu=10^{-5},~10^{-6}$). 
		The blue and green solid curves are $S(K)$ for the CH scheme with $\nu=10^{-6}$ and $\nu=10^{-5}$, respectively. 
		The red and orange dashed curves are $S(K)$ for the SH scheme with $\nu=10^{-6}$ and $\nu=10^{-5}$, respectively.}	
\label{fig:d=1e-5}
 \end{center}
\end{figure}

\subsection{Numerical Results}
The CHSH value $S$ in Eq.~(\ref{eq:CHSH}) is numerically calculated by using 
characteristic-function approach based on the covariance matrix of the quantum state and symplectic transformations~\cite{takeoka2015full,PhysRevA.93.042328,PhysRevA.98.063842}. 
See Supplemental Material and Ref.~\cite{takeoka2015full} for more details of this approach. 
Below we show the numerical results. 
When the corresponding fiber length 
is $L$~km, we set $\eta_{AH}=\eta_{AV}=\eta_T$ and $\eta_{BH}=\eta_{BV}=1$ for the SH scheme, and 
$\eta_{AH}=\eta_{AV}=\eta_{BH}=\eta_{BV}=\sqrt{\eta_T}$ for the CH scheme, respectively, where 
$\eta_T:=10^{-0.2L/10}$. 
In Fig.~\ref{fig:d=0}(a), we show the relation between $L$ and the CHSH value $S$ 
in an ideal system where all the local detection efficiencies are unity, the mode-matching is perfect, and detectors have no dark counts~(i.e. $\forall l$ $\eta_l$=1, $T_\mathrm{mode}=1$, and $\nu=0$ ). 
At each point, we perform the optimization over the average photon numbers of the TMSVs, and measurement angles using a random search algorithm. We see that the degradation of $S$ against the transmission distance is small for both of the CH and the SH schemes, 
since it is possible to set the optimal average photon numbers to be small~(typically $\sim10^{-5}$) in the ideal case. 
This makes the detrimental contribution of the multiple pairs negligible.  
Interestingly, the maximal violation at 0~km is $S\sim2.34$, 
which is slightly better than what is achieved by using a single-mode SPDC-based entangled pair source~(No ESR)~\cite{PhysRevA.98.063842,PhysRevA.91.012107}. 
On the other hand, the minimum detection efficiency to obtain $S>2$ is calculated to be $91.1\%$, 
which is larger than $66.7\%$ needed in the case of no ESR~\cite{PhysRevA.98.063842,PhysRevA.91.012107}. 
These differences come from the fact that the density operator of the heralded state is far from the state directly generated by SPDC which mainly consists of vaccum state.
The relation between $L$ and the key rate $K$ is shown in Fig.~\ref{fig:d=0}(b). 
The average photon numbers, which maximize $S$, are 
no longer optimal for maximizing $K$, since employing the small average photon numbers results in the low success probability $P^{\mathrm{suc}}$ of the Bell-state measurement at the ESR node. 
That is, there is a trade off between $S$ and $P^{\mathrm{suc}}$ for maximizing $K$. 
We clearly see the difference of $K$ between the CH scheme and the SH scheme. 
The reason is qualitatively understood as follows. 
In the SH scheme, since a large loss is imposed on the source~A, 
the average number of photons which survive at the ESR node is smaller 
than that in the CH scheme, which results in the lower $P^{\mathrm{suc}}$. 
Notice that $K$ in No ESR is much larger at 0~km, though $S$ is smaller than those using the ESR. 
This is because the violation of the CHSH inequality can occur with relatively large average photon numbers in No ESR, when a single-mode SPDC source is used~\cite{PhysRevA.98.063842}.  
Next, we add dark count probabilities of $\nu=10^{-6}$ and $\nu=10^{-5}$, and compare $S$ 
of the SH scheme and the CH scheme as shown in Fig.~\ref{fig:d=1e-5}(a). 
$S$ of the SH scheme starts to deviate from that of CH scheme for large $L$. 
The reason is also understood by the trade-off between $S$ and $P^{\mathrm{suc}}$.
When dark counts are considered, it is necessary to keep the average number of the photons that survive at ESR node sufficiently larger than the dark-count probability.
Thus, in the SH scheme, the optimal average photon number of the source~A must be 
larger than that in the CH scheme, which however results in smaller $S$. 
The minimum detection efficiencies to obtain $S>2$ slightly increase. 
For example, at $L=50$~km, $91.6\%$ and $92.7\%$ 
are necessary in the case of $\nu=10^{-6}$ and $\nu=10^{-5}$, respectively. 
We compare $K$ of the SH scheme and that of the CH scheme with considering 
the dark-count probabilities as shown in Fig.~\ref{fig:d=1e-5}~(b). We see a large gap between $K$ of the CH scheme and the SH scheme. 
When the total loss corresponds to the attenuation of the 100~km-optical fiber, 
$K$ of the CH scheme is about 100 times larger than that of the SH scheme 
that has been considered so far.

 \begin{figure*}[t]
 \begin{center}
\scalebox{0.42}{\includegraphics{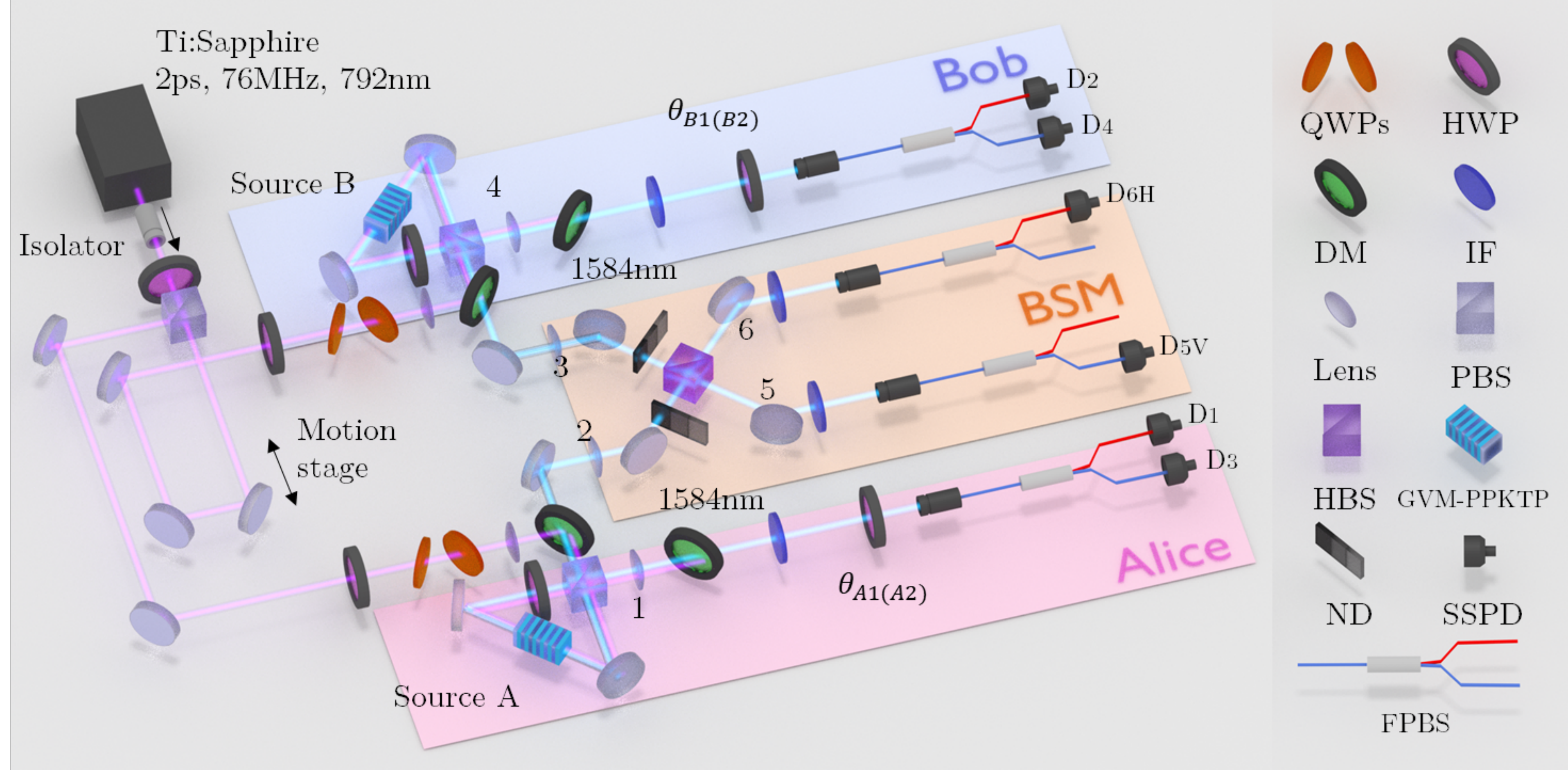}}
  \caption{The setup for the ESR-based Bell-test experiment. To generate entangled photon pairs by SPDC, 
we used counter propagating pump pulses to pump the GVM-PPKTP crystals in the Sagnac loop interferometers. 
Alice and Bob choose the measurement angles 
$\{\theta_{A1},\theta_{A2}\}$ and $\{\theta_{B1},\theta_{B2}\}$, respectively, 
and assign +1 or -1 for the each detection event to
calculate $S$ value.  GVM-PPKTP: group-velocity-matched periodically poled $\mathrm{KTiOPO_4}$, HBS: half beamsplitter, 
IF: interference filter, QWPs: paired quarter waveplates, HWP: half waveplate, DM: dichroic mirror, 
ND: neutral density filter, PBS: polarization beamsplitter, FPBS: 
fiber-based PBS, SSPD: superconducting single-photon detector.}
\label{fig:experiment}
 \end{center}
\end{figure*}

 \section{EXPERIMENT}
\label{secIII}
\subsection{Experimental Setup}
We perform the ESR-based Bell-test experiment using the setup illustrated in Fig.~\ref{fig:experiment}.
The pump pulse~(wavelength:~792~nm, pulse duration:~2~ps, repetition rate:~76~MHz) 
is obtained by a Ti:Sapphire laser.  
The pump pulse is split into two optical paths by a half waveplate~(HWP) and a polarization beamsplitter~(PBS), and fed to the 
two independent Sagnac-loop interferometers with group-velocity-matched periodically poled $\mathrm{KTiOPO_4}$~(GVM-PPKTP) crystals. 
The polarization of the each pump pulse is properly adjusted by a HWP and a paired quarter-waveplates~(QWPs). 
The two-qubit components of the states generated from the source~A and the source~B 
form the maximally entangled states 
$\ket{\Psi^{+}}_{12}$ and 
$\ket{\Psi^{-}}_{34}$, respectively. 
While the photon~1~(4) passes through the dichroic mirror~(DM) and goes to Alice's~(Bob's) side, the photons 2 and 3 
are led to the ESR node to perform the linear-optical BSM. 
The transmission losses in the optical fibers are emulated by two neutral density filters~(NDs) inserted in modes 2 and 3. 
In each optical path, we insert an interference filter~(IF) whose center wavelength and  bandwidth are 1584~nm and 2~nm, respectively, 
which is used to improve the purity of the SPDC photons. 
The linear optical BSM is implemented by mixing two input photons by means of a half beamsplitter~(HBS) 
followed by the polarization-dependent coincidence detection between $\mathrm{\mathrm{D_{5V}}}$ and $\mathrm{\mathrm{D_{6H}}}$, 
which projects the photon pair in modes 2 and 3 onto the singlet state $\ket{\Psi^-}_{23}$ with the success probability of 1/8. 
We note that if we introduce another two detectors and perform active feed forward, the maximum success probability becomes 1/2.
We use superconducting single-photon detectors~(SSPDs) whose quantum efficiency is $75~\%$ each~\cite{Miki:17}.
Alice and Bob set measurement angles $\{\theta_{A1},\theta_{A2}\}$ and $\{\theta_{B1},\theta_{B2}\}$, respectively,  
by means of the HWPs and fiber-based PBSs~(FPBSs). 
Finally, the photons are detected by four SSPDs: $\mathrm{\mathrm{D_1}}$ and $\mathrm{\mathrm{D_3}}$ for Alice, and $\mathrm{\mathrm{D_2}}$ and $\mathrm{\mathrm{D_4}}$ 
for Bob, respectively. 
In the experiment, the detection signal from $\mathrm{\mathrm{D_{5V}}}$ is used as a start signal for a time-to-digital converter (TDC), 
and the detection signals from $\mathrm{\mathrm{\mathrm{D_{6H}}}, \mathrm{D_1}, \mathrm{D_2}, \mathrm{D_3}}$ and $\mathrm{\mathrm{D_4}}$ are used as stop signals. 
Under the condition that the two-fold coincidence between $\mathrm{\mathrm{D_{5V}}}$ and $\mathrm{\mathrm{\mathrm{D_{6H}}}}$ occur, 
all the combination of click and no-click events are collected without postselection. 
We assign events of $\mathrm{\mathrm{D_1}~(\mathrm{D_2})}$ clicks on Alice's~(Bob's) side as $-1$ and all the others as $+1$, and then calculate $S$.

\subsection{Characterization of Experimental Setup}
\label{char}
We measure the experimental parameters which will be used in the numerical simulation. 
We first characterize the HBS at the ESR node using laser light centered at 1584~nm. 
It is found that the HBS is lossy only for the H-polarized light from mode~3. 
This loss is modeled by decomposing the HBS into the lossy material~($\eta_{AH}=0.27$) and 
the ideal HBS in the numerical simulation. 
We also characterize the local detection efficiencies $\eta_l$ for $l\in\{1,\cdots,6\}$ by using the weakly-pumped TMSVs~\cite{klyshko1980use}. 
The results are shown in Table~\ref{table:transimittance}. 
Throughout the experiment, we set the widths of the detection windows to be 1~ns. 
The dark-count rate within the detection window is measured to be $\nu=10^{-6}$. 

\begin{table}[htb]
\begin{center}
  \begin{tabular}{|c|c|c|} \hline
   $\eta_1$ & $\eta_2$ & $\eta_3$ \\ \hline 
   $14.63\pm2.75\%$ & $14.44\pm0.85\%$ & $10.87\pm2.36\%$ \\ \hline
   $\eta_4$ & $\eta_5$ & $\eta_6$ \\ \hline 
   $10.64\pm0.59\%$&$14.43\pm0.01\%$&$11.57\pm0.07\%$\\ \hline
  \end{tabular}
    \caption{The local detection efficiencies estimated by using  the weakly-pumped TMSV. }
\label{table:transimittance}
\end{center}
\end{table}

\begin{figure*}[t]
 \begin{center}
\scalebox{0.15}{\includegraphics{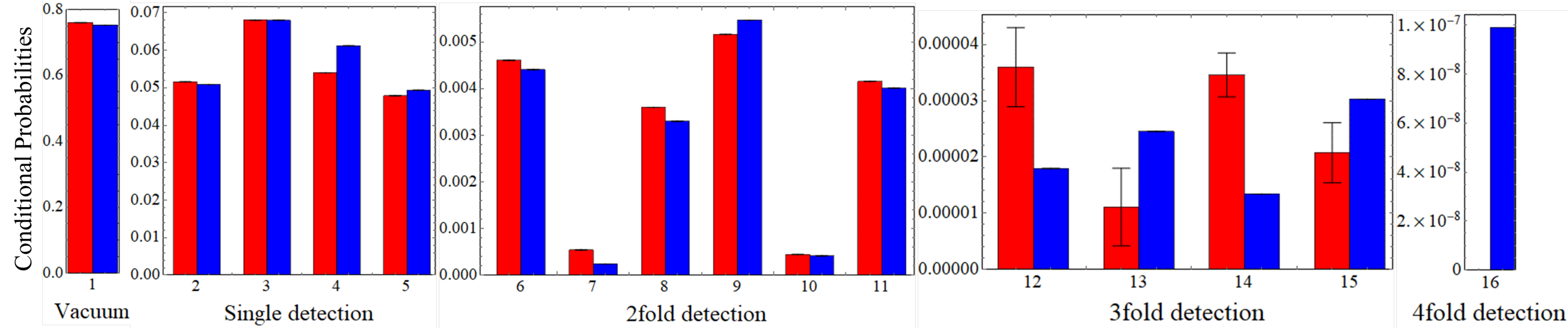}}
  \caption{The conditional detection probabilities obtained by the experiment~(red bars) and the numerical simulation~(blue bars). The error bars are 
	calculated by assuming the Poissonian distribution.
	Defining, for example, the conditional detection probability that only $\mathrm{\mathrm{D_1}}$ and $\mathrm{\mathrm{D_2}}$ fire 
	by $P(\mathrm{D_1}\cap \mathrm{D_2})$, the correspondence between the 16 labels and the 16 detection events are 
	described as follows: 1:$P(\mathrm{Vac})$, 2:$P(\mathrm{D_1})$, 3:$P(\mathrm{D_2})$, 4:$P(\mathrm{D_3})$, 5:$P(\mathrm{D_4})$, 6:$P(\mathrm{D_1}\cap \mathrm{D_2})$, 7:$P(\mathrm{D_1}\cap \mathrm{D_3})$, 
	8:$P(\mathrm{D_1}\cap \mathrm{D_4})$, 9:$P(\mathrm{D_2}\cap \mathrm{D_3})$, 10:$P(\mathrm{D_2}\cap \mathrm{D_4})$, 11:$P(\mathrm{D_3}\cap \mathrm{D_4})$, 12:$P(\mathrm{D_1}\cap \mathrm{D_2} \cap \mathrm{D_3})$, 13:$P(\mathrm{D_1}\cap \mathrm{D_2}\cap \mathrm{D_4})$, 
	14:$P(\mathrm{D_1}\cap \mathrm{D_3}\cap \mathrm{D_4})$, 15:$P(\mathrm{D_2}\cap \mathrm{D_3}\cap \mathrm{D_4})$, and 16:$P(\mathrm{D_1}\cap \mathrm{D_2}\cap \mathrm{D_3}\cap \mathrm{D_4})$, 
	where $P(\mathrm{Vac})$ is the conditional probability that none of $\mathrm{\mathrm{D_1}, \mathrm{D_2}, \mathrm{D_3}}$ and $\mathrm{\mathrm{D_4}}$ fires. }
\label{fig:prob}
 \end{center}
\end{figure*}

\begin{table}[htb]
\begin{center}
  \begin{tabular}{|c|c|c|c|c|} \hline
  & $\mu_1$ & $\mu_2$ & $\mu_3$ & $\mu_4$ \\ \hline 
   Optimal & $3.95\times10^{-2}$ & $1.50\times10^{-2}$ & $1.50\times10^{-2}$  & $1.50\times10^{-2}$\\ \hline
   Experiment  & $3.83\times10^{-2}$ & $1.48\times10^{-2}$ & $1.64\times10^{-2}$  & $1.52\times10^{-2}$\\ \hline
  \end{tabular}
    \caption{The optimal average photon numbers of the TMSVs, and the average photon numbers estimated by the experiment. }
\label{table:average}
\end{center}
\end{table}

Under the above experimental conditions, we perform the numerical optimization of 
the average photon numbers of the TMSVs and the measurement angles
such that $S$ is maximized. 
Note that, in the optimization, we assume $\eta_1=\eta_2=\eta_3=\eta_4=1$, 
since the detection efficiencies shown in Table~\ref{table:transimittance} are not sufficient to 
observe the detection-loophole-free violation of the CHSH inequality. 
In addition, we impose a condition that each average photon number is at least $\geq1.5\times10^{-2}$ to 
finish the experiment within reasonable time.
We set the average photon numbers of the TMSVs based on the 
numerical results. 
The optimal average photon numbers and the experimentally-measured ones are shown in Table~\ref{table:average}, where 
$\mu_k$ is the average photon number of TMSV$k$. 
We see that $\mu_1$ is larger than the others, 
since $\eta_{AH}$ is imposed in the transmission path of $\mathrm{TMSV}1$. 
The optimal measurement angles are $\{\theta_{A1}, \theta_{A2}\}=\{0,~0.58\}$~[rad] and  $\{\theta_{B1},\theta_{B2}\}=\{1.47,~2.01\}$~[rad]. 
With above experimental parameters, 
the two-qubit components of the input quantum states and the indistinguishability between photon~3 and the photon~4 are characterized.~(see Supplemental Material.)

\subsection{Experimental Results}
We adopt the CH scheme, and perform the ESR-based Bell experiment. 
Under the condition of the successful BSM, we accumulate 
every detection event of the heralded state without postselection. 
First, we remove the ND filters, and perform the Bell-test experiment on the heralded state with the optimal measurement angles. 
Since the detection efficiencies of our system are not in the range of 
closing the detection loophole, $S$ does not directly exceed the threshold value of $S=2$. 
In fact, when we input all the experimental parameters to the numerical simulation, 
the value of $S$ is expected to be $S_{\mathrm{th}}=1.486$. 
Nevertheless, it is still possible to compare $S_{\mathrm{th}}$ and the CHSH value obtained by the experiment~$S_{\mathrm{exp}}$. 
From the experimentally-obtained conditional probability distributions, $S_{\mathrm{exp}}$ is calculated to be $S_{\mathrm{exp}}=1.481\pm0.002$, 
which coincides with $S_{\mathrm{th}}$. 
We also compare the conditional detection probabilities.  
For example, all the conditional detection probabilities for $\{\theta_{A1},\theta_{B1}\}=\{0,~1.47\}$~[rad] are shown in Fig.~\ref{fig:prob}. 
Since each of Alice and Bob possesses two detectors, there are $2^{4}=16$ possible detection events for each measurement angle. 
The red bars and blue bars correspond to the conditional probabilities obtained by the experiment and 
the numerical simulation, respectively. 
We clearly see an excellent agreement between the experimental results and the numerical simulations. 
Moreover, the L1-distance $\mathcal{D}$ defined by $\mathcal{D}=\sum_{i=1}^{16}|p_i-q_i|$ is calculated to be 
as small as $\mathcal{D}=0.018\pm0.001$, 
where, $p_i$($q_i$) is the $i$-th experimentally~(theoretically)-obtained conditional detection probability, respectively. 
For the other measurement angles, the L1-distances are calculated to be $\mathcal{D}=0.020\pm0.001$ for $\{\theta_{A1},\theta_{B2}\}=\{0,~2.01\}$~[rad], 
$\mathcal{D}=0.029\pm0.001$ for $\{\theta_{A2},\theta_{B1}\}=\{0.58,~1.47\}$~[rad] and $\mathcal{D}=0.017\pm0.001$ for $\{\theta_{A2},\theta_{B2}\}=\{0.58,~2.01\}$~[rad]. 

\begin{figure}[t]
 \begin{center}
\scalebox{0.25}{\includegraphics{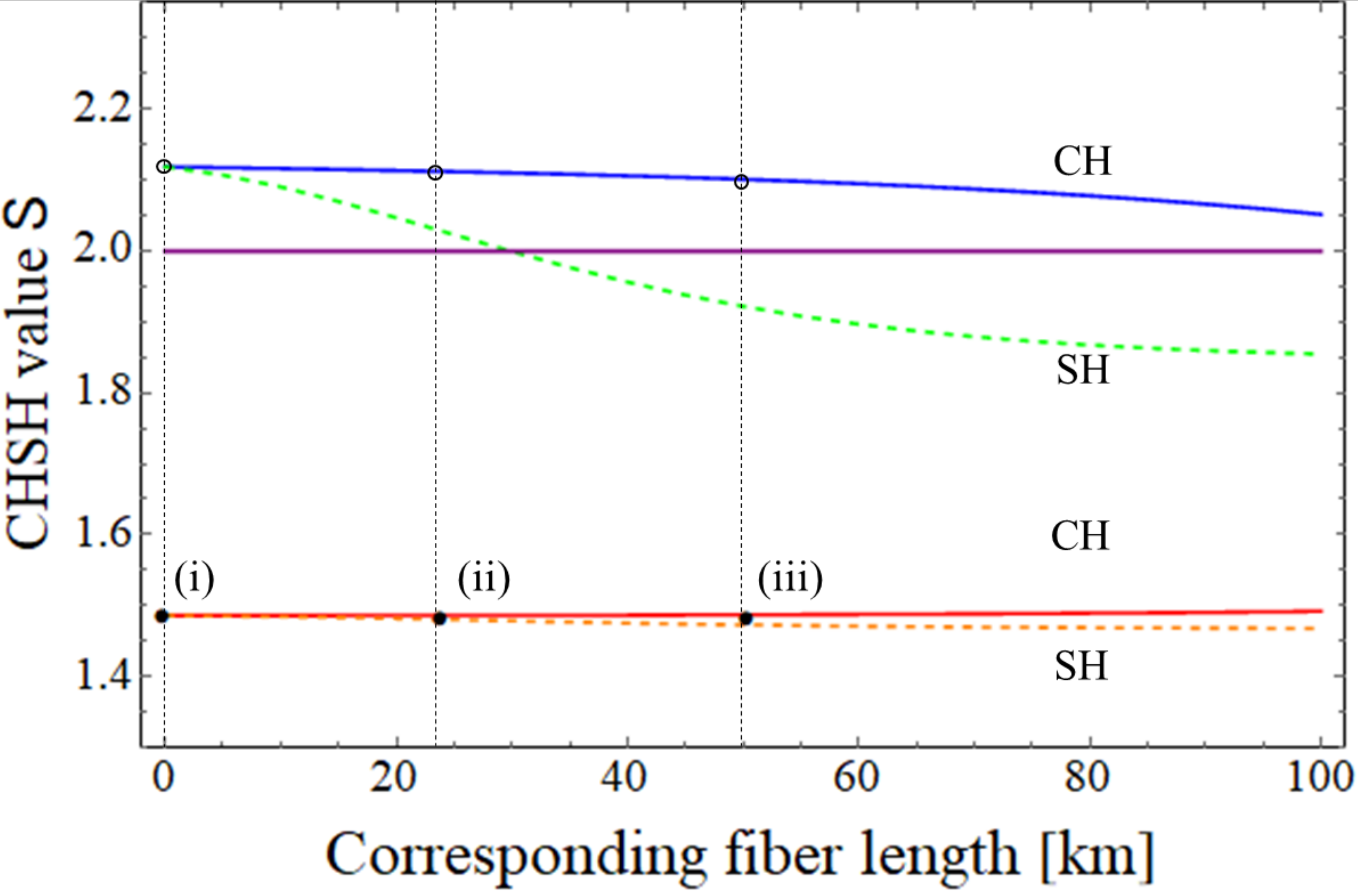}}
  \caption{The corresponding fiber length $L$ vs $S$. 
	The red solid curve~(the CH scheme) and the orange dashed curve~(the SH scheme) are obtained by the numerical simulation with 
	all of the experimental imperfections. 
	The blue solid curve~(the CH scheme) and green dashed curve~(the SH scheme) are obtained by the numerical simulation with assuming that $\eta_l=1$ for $l\in\{1,2,3,4\}$. 
	The black dots are $S$ obtained by the experiment. 
	The circles on the blue solid curve are $S$ of the heralded states just before detection. The purple solid line is the threshold value of $S=2$. 
	 }	
\label{fig:DvsS}
 \end{center}
\end{figure}

Next, we insert the ND filters, and perform the Bell-test experiment on the heralded state while changing the transmission losses. 
Note that we fix the average photon numbers and measurement angles throughout the experiment. 
The results are shown in Fig.~\ref{fig:DvsS} as three black dots. 
The total transmittance of the ND filters are equivalent to (i)~0~km, (ii)~24~km and (iii)~50~km of the optical fibers, 
and the corresponding $S$ are (i)~$S_{\mathrm{exp}}=1.481\pm0.002$, (ii)~$S_{\mathrm{exp}}=1.479\pm0.002$ and (iii)~$S_{\mathrm{exp}}=1.478\pm0.002$. 
They agree well with the theoretical curve for the CH scheme~(shown by a red solid curve) obtained by using the experimental parameters 
characterized by a separate measurement in Sec.~\ref{char}. 
When the detection efficiencies are small, the difference between the CH scheme and SH scheme~(shown by a orange dashed curve) is small. 
The blue solid curve~(the CH scheme) and green dashed curve~(the SH scheme) are obtained by the numerical simulation with 
the experimental parameters but 
assuming that $\eta_l=1$ for $l\in\{1,2,3,4\}$. 
Since our model fits the experimental results, 
these curves are considered to be the nonlocality of the heralded state just before detection. 
Interestingly, there is a large gap between the CH scheme and the SH scheme. 
The estimated CHSH values~($S^{\eta=1}$) are shown by the three circles in Fig.~\ref{fig:DvsS}. 
The values are estimated to be $S^{\eta=1}$=2.120~(0~km),  2.115~(24~km) and 2.104~(50~km), respectively, 
which indicates that the quantum state just before detection possesses potential to violate the CHSH inequality 
even with various experimental imperfections.

\section{DISCUSSION}
\label{secIV}
 \begin{figure}[t]
 \begin{center}
\scalebox{0.27}{\includegraphics{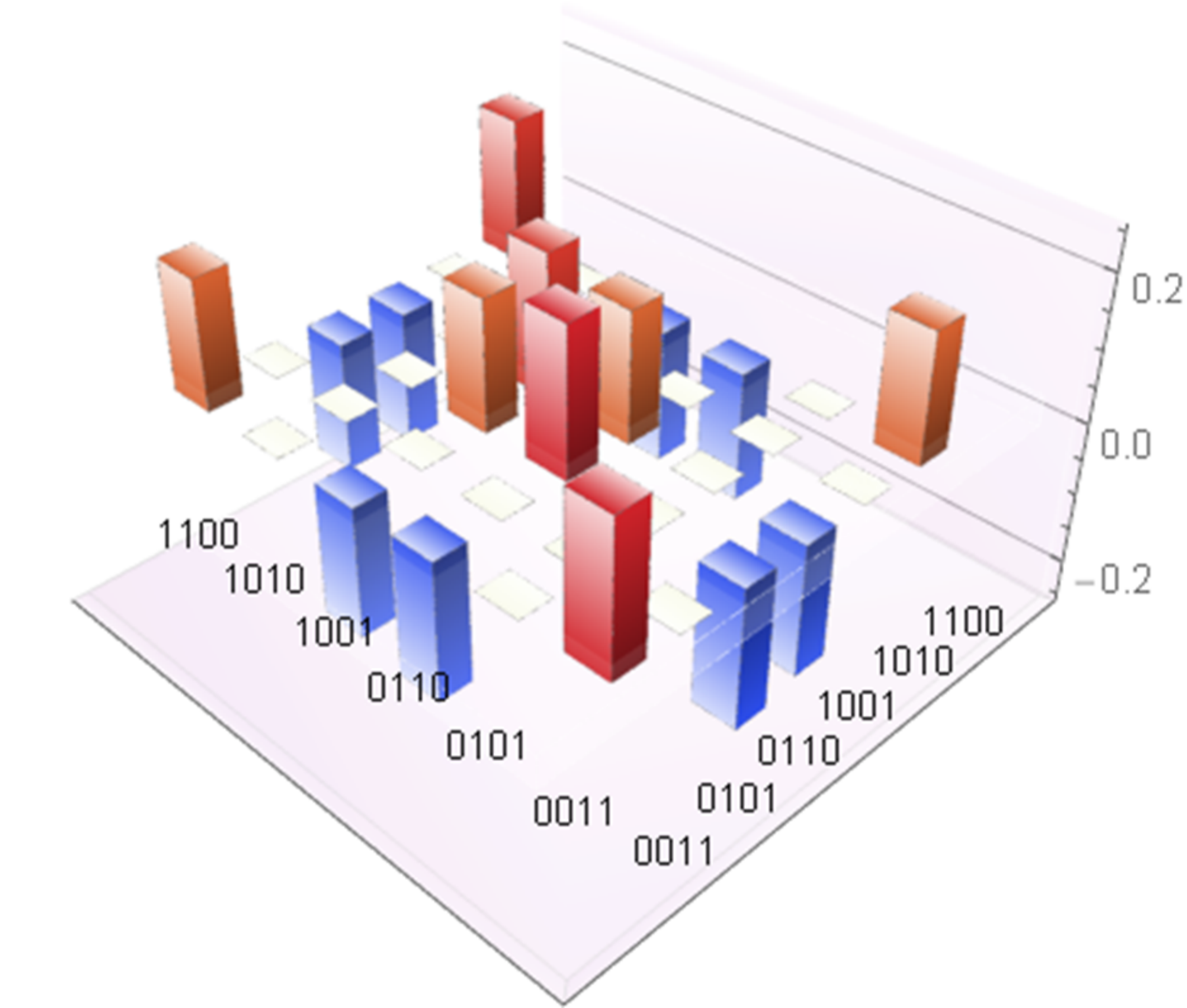}}
  \caption{
The real part of the partial density matrix 
of the heralded state spanned by $\ket{0011}$, $\ket{0101}$, $\ket{0110}$, $\ket{1001}$, $\ket{1010}$, and 
$\ket{0011}$. The matrix elements of the imaginary part are almost zero. }
\label{fig:swapdens}
 \end{center}
\end{figure}
In this section, we estimate the density matrix of the experimentally heralded state just before detection 
by compensating the detection inefficiency with the help of our theoretical model. 
As shown in Fig.~\ref{fig:ESRREAL}, the heralded state is distributed over the four modes: 
$H_1$, $V_1$, $H_4$, and $V_4$. 
In addition, as described in Sec.~\ref{secI}, the successful BSM mainly consists of the superposition of the following four events 
(i)~one photon in each of mode $H_1$, $V_2$, $V_1$ and $H_2$, (ii)~one photon in each of mode $H_3$, $V_4$, $V_3$ and $H_4$, 
(iii)~one photon in each of mode $H_1$, $V_2$, $H_3$ and $V_4$, and (iv)~one photon in each of mode $V_1$, $H_2$, $V_3$ and $H_4$. 
Thus, we restrict ourselves to the subspace spanned by 
$\{\ket{0011}, \ket{0101},\ket{0110},\ket{1001},\ket{1010},
\ket{1100}\}$, 
where the modes are arranged in order of $H_1$, $V_1$, $H_4$, and $V_4$.
By the numerical simulation, we know the characteristic function of the heralded state $\chi_{\hat{\rho}^{\mathrm{herald}}_{H_1V_1H_4V_4}}(\xi)$.~(The explicit formula is 
given in Supplemental Material.)  
Thus, the matrix elements of $\hat{\rho}^{\mathrm{herald}}_{H_1V_1H_4V_4}$ in the Fock-state basis are 
calculated by the inner product of $\chi_{\hat{\rho}^{\mathrm{herald}}_{H_1V_1H_4V_4}}(\xi)$ and the 
characteristic function of the corresponding four-mode Fock state. 
For example, $\bra{0011}\hat{\rho}\ket{1100}$ is calculated by 
\begin{equation}
\left(\frac{1}{2\pi}\right)^4\int \chi_{\hat{\rho}^{\mathrm{herald}}_{H_1V_1H_4V_4}}(\xi)\chi_{\ketbra{1100}{0011}}(-\xi)d\xi,
\label{eq:fockdens}
\end{equation}
where $\chi_{\ketbra{1100}{0011}}(\xi)$ is the characteristic function of $\ketbra{1100}{0011}$. 
We use the characteristic function of the heralded state for 50~km, 
and reconstruct the unnormalized partial density matrix, as shown in Fig.~\ref{fig:swapdens}.~(see Supplemental Material for the detailed calculation.) 
In addition to the four center peaks which correspond to $\ketbra{\Psi^+}{\Psi^+}$, 
we clearly see the contribution of the events (i) and (ii). 
By renormalizing this partial density matrix, 
the fidelity to $\ketbra{\Psi^+}{\Psi^+}$
is calculated to be 0.47. This indicates that the heralded state is 
clearly far from the two-qubit maximally entangled states, while it possesses enough nonlocality to violate the CHSH inequality.

\section{CONCLUSION}
\label{secV}
In conclusion, 
we propose and demonstrate 
an improved version of the heralded nonlocality amplifier based on the ESR. 
In theory, we employ the method to calculate the detection probabilities using the characteristic function, 
and reveal that, for both of the key rate and the CHSH value $S$,
the improved scheme~(CH scheme: the ESR node is placed in the middle of Alice and Bob) 
is much more robust against the transmission loss than 
the previous one~(SH scheme: the ESR node is placed in Bob's side). 
Importantly, the larger the dark count rate is, the larger the gap of the performance between the CH scheme and the SH scheme becomes. 
In experiment, we perform the ESR-based Bell-test 
using the optimal parameters derived by the numerical simulation. 
While the detection efficiencies of our system is not in the range of closing 
the detection loophole, the experimental results are 
in excellent agreement with 
the numerical simulation with experimental parameters which are characterized in a separate measurement. 
This allows us to estimate the nonlocality and the density matrix of the heralded state just before detection. 
It is revealed that, while the density matrix of the heralded state is far from the ideal two-qubit maximally entangled state, 
the state 
possesses strong nonlocality~($S^{\eta=1}=2.104>2$) 
after the transmission loss of 10~dB which is equivalent to 50~km-long optical fibers at telecommunication wavelength. 
To directly observe $S > 2$ over 50~km, it is found that a detection efficiency at least $97.4\%$ is necessary with our current experimental conditions.
However, the threshold detection efficiency can be improved further down to $91.6\%$, if the experimental imperfections other than the dark counts are reduced. 
In view of the recent progress of the single-photon detection highlighted  by high-efficiency single-photon detectors with quantum efficiencies $>93~\%$~\cite{marsili2013detecting,Lita:08}, 
it could be possible to experimentally observe the nonlocality over such a long distance. 
Our result thus shows an important benchmark about the ESR protocol, and represents a major building block towards the long-distance realization of the loophole-free test
of the CHSH-violation as well as DIQKD.

\section{Supplementary Material}
\subsection{Detailed Calculations Based
 On The Characteristic Function}
In this section, we present the detailed method to compute 
the conditional detection probabilities 
using the theoretical model in Fig.~\ref{fig:ESRREAL}. 
We follow the definitions introduced in Ref.~\cite{PhysRevA.98.063842}. 
We define a density operator acting on the N-dimensional Hilbert space $\mathcal{H}^{\otimes N}$ as $\hat{\rho}$. 
The characteristic function of $\hat{\rho}$ is defined by 
\begin{equation}
\chi(\xi)=\mathrm{Tr}[\hat{\rho}\hat{\mathcal{W}}(\xi)],
\end{equation}
where 
\begin{equation}
\hat{\mathcal{W}}(\xi)=\mathrm{exp}\left(-i\xi^T\hat{R}\right)
\end{equation}
is the Weyl operator. Here, 
$\hat{R}=(\hat{x}_1,\dots,\hat{x}_N,\hat{p}_1,\dots,\hat{p}_N)$ and 
$\xi=(\xi_1,\dots,\xi_{2N})$ 
are a 2$N$ vector consisting of quadrature operators and a 2$N$ real vector, respectively. 
When the characteristic function of the quantum state has 
a Gaussian distribution 
\begin{equation}
\chi(\xi)=\mathrm{exp}\left(-\frac{1}{4}\xi^T\gamma\xi-id^T\xi\right), 
\end{equation}
the quantum state is simply characterized 
by a 2$N$ $\times$ 2$N$ matrix $\gamma$~(the covariance matrix) 
and a 2$N$-dimensional vector $d$~(the displacement vector). 

\begin{figure*}[t]
 \begin{center}
\scalebox{0.37}{\includegraphics{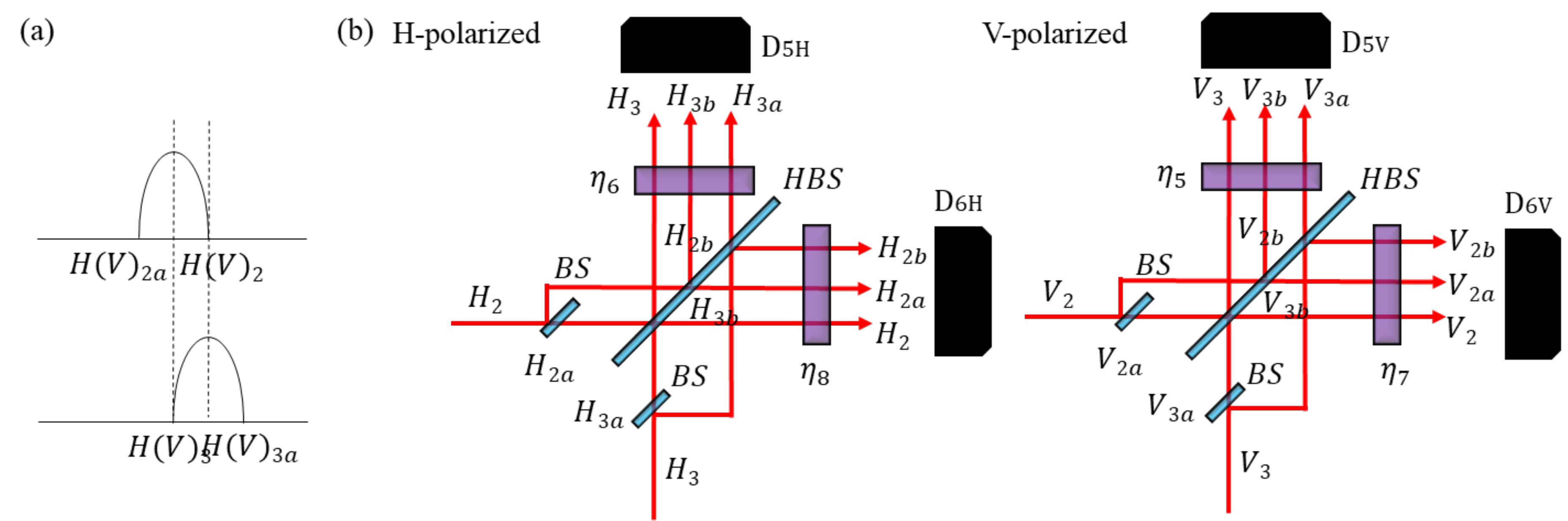}}
  \caption{(a)~The sketch of the mode mismatch. Each of the light pulse is 
  divided into two fractions: the fraction which interferes with probability amplitude $\sqrt{T_\mathrm{mode}}$ and 
  the fraction which does not interfere with probability amplitude $\sqrt{1-T_\mathrm{mode}}$. (b)~The model of the mode mismatch. 
  The virtual BSs with transmittance $T_\mathrm{mode}$ are inserted before the HBS.}
\label{fig:modematching}
 \end{center}
\end{figure*}

In our theoretical model, each entangled photon pair source consists of 
two TMSV sources over polarization modes embedded in the Sagnac loop. 
The covariance matrices of the quantum state from source~A~($\gamma_{{H_1V_1H_2V_2}}^{\mathrm{SA}}$) and source~B~($\gamma_{{H_3V_3H_4V_4}}^{\mathrm{SB}}$) are given by Refs.~\cite{takeoka2015full,PhysRevA.98.063842} 
\begin{equation}
\gamma_{H_1V_1H_2V_2}^{\mathrm{SA}}=\left[
	\begin{array}{cc}
	\gamma^{\mathrm{SA1}}(\mu_1,\mu_2)&\mathbf{0}\\
	\mathbf{0}&\gamma^{\mathrm{SA2}}(\mu_1,\mu_2)
	\end{array}
	\right]
\label{TMSVSLA}
\end{equation}
and
\begin{equation}
\gamma_{H_1V_1H_2V_2}^{\mathrm{SB}}=\left[
	\begin{array}{cc}
	\gamma^{\mathrm{SB1}}(\mu_3,\mu_4)&\mathbf{0}\\
	\mathbf{0}&\gamma^{\mathrm{SB2}}(\mu_3,\mu_4)
	\end{array}
	\right], 
\label{TMSVSLB}
\end{equation}
respectively, where 
\begin{widetext}
\begin{align}
&\gamma^{\mathrm{SA1}}(\mu_1,\mu_2)=\left[
	\begin{array}{cccc}
	2\mu_{1}+1&0&0&2\sqrt{\mu_{1}(\mu_{1}+1)}\\
	0&2\mu_{2}+1&2\sqrt{\mu_{2}(\mu_{2}+1)}&0\\
	0&2\sqrt{\mu_{2}(\mu_{2}+1)}&2\mu_{2}+1&0\\
	2\sqrt{\mu_{1}(\mu_{1}+1)}&0&0&2\mu_{1}+1
	\end{array}
	\right], 
\label{TMSVSLA1}\\
&\gamma^{\mathrm{SA2}}(\mu_1,\mu_2)=\left[
	\begin{array}{cccc}
	2\mu_{1}+1&0&0&-2\sqrt{\mu_{1}(\mu_{1}+1)}\\
	0&2\mu_{2}+1&-2\sqrt{\mu_{2}(\mu_{2}+1)}&0\\
	0&-2\sqrt{\mu_{2}(\mu_{2}+1)}&2\mu_{2}+1&0\\
	-2\sqrt{\mu_{1}(\mu_{1}+1)}&0&0&2\mu_{1}+1
	\end{array}
	\right], 
\label{TMSVSLA2}\\
&\gamma^{\mathrm{SB1}}(\mu_3,\mu_4)=\left[
	\begin{array}{cccc}
	2\mu_{3}+1&0&0&2\sqrt{\mu_{3}(\mu_{3}+1)}\\
	0&2\mu_{4}+1&-2\sqrt{\mu_{4}(\mu_{4}+1)}&0\\
	0&-2\sqrt{\mu_{4}(\mu_{4}+1)}&2\mu_{4}+1&0\\
	2\sqrt{\mu_{3}(\mu_{3}+1)}&0&0&2\mu_{3}+1
	\end{array}
	\right], 
\label{TMSVSLB1}\\
&\gamma^{\mathrm{SB2}}(\mu_3,\mu_4)=\left[
	\begin{array}{cccc}
	2\mu_{3}+1&0&0&-2\sqrt{\mu_{3}(\mu_{3}+1)}\\
	0&2\mu_{4}+1&2\sqrt{\mu_{4}(\mu_{4}+1)}&0\\
	0&2\sqrt{\mu_{4}(\mu_{4}+1)}&2\mu_{4}+1&0\\
	-2\sqrt{\mu_{3}(\mu_{3}+1)}&0&0&2\mu_{3}+1
	\end{array}
	\right]. 
\label{TMSVSLB2}
\end{align}
\end{widetext}
The overall input quantum state is described by 
$\gamma^{\mathrm{in}}_{\mathcal{S}}:=\gamma_{{H_1V_1H_2V_2}}^{\mathrm{SA}}\oplus\gamma_{{H_3V_3H_4V_4}}^{\mathrm{SB}}$, 
where $\mathcal{S}:=\{H_1,V_1,H_2,V_2,H_3,V_3,H_4,V_4\}$. 
The photons in modes $\mathrm{H_2, V_2, H_3}$ and $\mathrm{V_3}$ are sent to the ESR node through the transmission losses. 
We describe the transformation of the linear loss with transmittance $t$ on a single-mode Gaussian state with covariance matrix $\gamma$ by 
\begin{equation}
    \mathcal{L}^t\gamma=K^T\gamma K+\alpha,
\end{equation}
where $K=\sqrt{t}I$ and $\alpha=(1-t)I$. 
Then, the linear losses $\eta_{AH}$, $\eta_{AV}$, $\eta_{BH}$ and $\eta_{BV}$ transform the input covariance matrix~$\gamma^{\mathrm{in}}_{\mathcal{S}}$ into 
\begin{widetext}
\begin{eqnarray}
\gamma^{\mathrm{Loss}}_{\mathcal{S}}&=&\mathcal{L}^{\eta_{AH}}_{H_2}\mathcal{L}^{\eta_{AV}}_{V_2}\mathcal{L}^{\eta_{BH}}_{H_3}\mathcal{L}^{\eta_{BV}}_{V_3} \gamma^{\mathrm{in}}_{\mathcal{S}}\\
&=&\left(K^{\eta_{AH}\eta_{AV}\eta_{BH}\eta_{BV}}_{H_2V_2H_3V_3}\right)^T \gamma^{\mathrm{in}}_{\mathcal{S}}K^{\eta_{AH}\eta_{AV}\eta_{BH}\eta_{BV}}_{H_2V_2H_3V_3}
+\alpha^{\eta_{AH}\eta_{AV}\eta_{BH}\eta_{BV}}_{H_2V_2H_3V_3}, 
\end{eqnarray}
\end{widetext}
where 
\begin{widetext}
\begin{equation}
K^{\eta_{AH}\eta_{AV}\eta_{BH}\eta_{BV}}_{H_2V_2H_3V_3}=	
	\left[
	\begin{array}{cccccccc}
	1&0&0&0&0&0&0&0\\
	0&1&0&0&0&0&0&0\\
	0&0&\sqrt{\eta_{AH}}&0&0&0&0&0\\
	0&0&0&\sqrt{\eta_{AV}}&0&0&0&0\\
	0&0&0&0&\sqrt{\eta_{BH}}&0&0&0\\
	0&0&0&0&0&\sqrt{\eta_{BV}}&0&0\\
	0&0&0&0&0&0&1&0\\
	0&0&0&0&0&0&0&1
	\end{array}
	\right]^{\oplus2}. 
\end{equation}
\end{widetext}
and
\begin{widetext}
\begin{equation}
\alpha^{\eta_{AH}\eta_{AV}\eta_{BH}\eta_{BV}}_{H_2V_2H_3V_3}=	
	\left[
	\begin{array}{cccccccc}
	1&0&0&0&0&0&0&0\\
	0&1&0&0&0&0&0&0\\
	0&0&1-\eta_{AH}&0&0&0&0&0\\
	0&0&0&1-\eta_{AV}&0&0&0&0\\
	0&0&0&0&1-\eta_{BH}&0&0&0\\
	0&0&0&0&0&1-\eta_{BV}&0&0\\
	0&0&0&0&0&0&1&0\\
	0&0&0&0&0&0&0&1
	\end{array}
	\right]^{\oplus2}. 
\end{equation}
\end{widetext}
Here, for simplicity, we represent the block diagonal matrix like $\left[
	\begin{array}{cc}
	A&\mathbf{0}\\
	\mathbf{0}&A
	\end{array}
	\right]$
by $A^{\oplus2}$.
As described in Sec.~\ref{secII}, the mode matching between photon~($H_2\&H_3$) and ($V_2\&V_3$) 
are considered by dividing the each input light pulse into two mutually orthogonal modes
as shown in Fig.~\ref{fig:modematching}(a). 
This is modeled by inserting virtual BSs whose transmittance are $T_\mathrm{mode}$ before the HBS as shown in Fig~\ref{fig:modematching}(b). 
The fractions with probability $T_\mathrm{mode}$ interfere at the HBS, while the fractions with probability $1-T_\mathrm{mode}$ are mixed with vacua by the HBS. 
In the numerical simulation, we first add the eight modes~($H(V)_{2a}$, $H(V)_{3a}$, $H(V)_{2b}$ and $H(V)_{3b}$) of vacua to $\gamma^{\mathrm{Loss}}_{\mathcal{S}}$ as 
$\gamma^{\mathrm{MM}}_{\mathcal{U}}:=\gamma^{\mathrm{Loss}}_{\mathcal{S}}\oplus I_{H_{2a}\ldots H_{3b}V_{2a}\ldots V_{3b}}$, 
where $\mathcal{U}:=\mathcal{S}\cup\{H_{2a}\ldots H_{3b}V_{2a}\ldots V_{3b}\}$. 
Second, we perform the symplectic transformations of the BSs as 
\begin{widetext}
\begin{equation}
    \gamma^{\mathrm{BS}}_{\mathcal{U}}=(S^{\theta_{T_\mathrm{mode}}}_{H_2H_{2a}}\oplus S^{\theta_{T_\mathrm{mode}}}_{H_3H_{3a}}\oplus S^{\theta_{T_\mathrm{mode}}}_{V_2V_{2a}}\oplus S^{\theta_{T_\mathrm{mode}}}_{V_3V_{3a}})^T\gamma^{\mathrm{MM}}_{\mathcal{U}} (S^{\theta_{T_\mathrm{mode}}}_{H_2H_{2a}}\oplus S^{\theta_{T_\mathrm{mode}}}_{H_3H_{3a}}\oplus S^{\theta_{T_\mathrm{mode}}}_{V_2V_{2a}}\oplus S^{\theta_{T_\mathrm{mode}}}_{V_3V_{3a}}), 
\end{equation}
\end{widetext}
where $\theta_{T_\mathrm{mode}}:=\arccos{\sqrt{T_\mathrm{mode}}}$, and 
\begin{equation}
S^{\theta}_{ij}:=	
	\left[
	\begin{array}{cc}
	\cos{\theta}&\sin{\theta}\\
	-\sin{\theta}&\cos{\theta}
	\end{array}
	\right]^{\oplus2} 
\end{equation}
is the symplectic matrix of the BS whose transmittance is $\cos^2{\theta}$
acting on the modes $i$ and $j$. 
Finally, we perform the symplelctic transformation of the HBSs as 
\begin{widetext}
\begin{eqnarray}
    \gamma^{\mathrm{HBS}}_{\mathcal{U}}=(S^{\pi/4}_{H_2H_3}\oplus S^{\pi/4}_{H_{2a}H_{3b}}\oplus S^{\pi/4}_{H_{3a}H_{2b}}\oplus S^{\pi/4}_{V_2V_3}\oplus S^{\pi/4}_{V_{2a}V_{3b}}\oplus S^{\pi/4}_{V_{3a}V_{2b}})^T\nonumber\\
    \gamma^{\mathrm{BS}}_{\mathcal{U}}(S^{\pi/4}_{H_2H_3}\oplus S^{\pi/4}_{H_{2a}H_{3b}}\oplus S^{\pi/4}_{H_{3a}H_{2b}}\oplus S^{\pi/4}_{V_2V_3}\oplus S^{\pi/4}_{V_{2a}V_{3b}}\oplus S^{\pi/4}_{V_{3a}V_{2b}}).  
\end{eqnarray}
\end{widetext}

We consider the imperfect detection efficiency of each heralding detector at the ESR node as 
\begin{widetext}
\begin{equation}
\gamma^{\mathrm{BSM}}_{\mathcal{U}}=\mathcal{L}^{\eta_8}_{H_2}\mathcal{L}^{\eta_8}_{H_{2a}}\mathcal{L}^{\eta_8}_{H_{2b}}
\mathcal{L}^{\eta_7}_{V_2}\mathcal{L}^{\eta_7}_{V_{2a}}\mathcal{L}^{\eta_7}_{V_{2b}}
\mathcal{L}^{\eta_6}_{H_3}\mathcal{L}^{\eta_6}_{H_{3a}}\mathcal{L}^{\eta_6}_{H_{3b}}
\mathcal{L}^{\eta_5}_{V_3}\mathcal{L}^{\eta_5}_{V_{3a}}\mathcal{L}^{\eta_5}_{V_{3b}}
\gamma^{\mathrm{HBS}}_{\mathcal{U}}. 
\end{equation}
\end{widetext}
The successful BSM corresponds to the two-fold coincidence between $(\mathrm{D_{5H}}\cap \mathrm{D_{6V}})$ or $(\mathrm{D_{5V}}\cap \mathrm{\mathrm{D_{6H}}})$. For example, the two-fold coincidence probability $P(\mathrm{D_{5V}}\cap \mathrm{\mathrm{D_{6H}}})$ is given by 
\begin{widetext}
\begin{align}
&P(\mathrm{D_{5V}}\cap \mathrm{\mathrm{D_{6H}}})=\mathrm{Tr}[\hat{\rho}^{\gamma^{\mathrm{BSM}}_{\mathcal{U}}}(\hat{I}-\hat{\Pi}^{\mathrm{off}}_{V_3}\hat{\Pi}^{\mathrm{off}}_{V_{3a}}\hat{\Pi}^{\mathrm{off}}_{V_{3b}})(\hat{I}-\hat{\Pi}^{\mathrm{off}}_{H_2}\hat{\Pi}^{\mathrm{off}}_{H_{2a}}\hat{\Pi}^{\mathrm{off}}_{H_{2b}})]\label{Eq:on-off}\\
&=\mathrm{Tr}[\rho^{\gamma^{\mathrm{BSM}}_{\mathcal{U}}}(\hat{I}-(1-\nu)^3\ketbra{0}{0}^{\otimes3}_{V_3V_{3a}V_{3b}})(\hat{I}-(1-\nu)^3\ketbra{0}{0}^{\otimes3}_{H_2H_{2a}H_{2b}})]\\
&=\mathrm{Tr}[\rho^{\gamma^{\mathrm{BSM}}_{\mathcal{U}}}(\hat{I}-(1-\nu)^3\ketbra{0}{0}^{\otimes3}_{V_3V_{3a}V_{3b}}\nonumber\\
&+(1-\nu)^6\ketbra{0}{0}^{\otimes6}_{V_3V_{3a}V_{3b}H_2H_{2a}H_{2b}}-(1-\nu)^3\ketbra{0}{0}^{\otimes3}_{H_2H_{2a}H_{2b}})]\\
&=1-\frac{8(1-\nu)^3}{\sqrt{\mathrm{det}(\gamma_{V_3V_{3a}V_{3b}}^{\mathrm{BSM}}+I)}}+\frac{64(1-\nu)^6}{\sqrt{\mathrm{det}(\gamma_{V_3V_{3a}V_{3b}H_2H_{2a}H_{2b}}^{\mathrm{BSM}}+I)}}-\frac{8(1-\nu)^3}{\sqrt{\mathrm{det}(\gamma_{H_2H_{2a}H_{2b}}^{\mathrm{BSM}}+I)}}
,\label{Eq:Psuc}\\
&=:P_0-P_1+P_2-P_3
\end{align}
\end{widetext}
where $\gamma_{j_1\ldots j_n}^{\mathrm{BSM}}$ is the submatrix 
obtained by extracting the rows and columns corresponding to modes $j_1\ldots j_n$ 
from $\gamma^{\mathrm{BSM}}_{\mathcal{U}}$. 
In Eq.~(\ref{Eq:on-off}), we use 
the POVM elements of 
the on-off detector acting in mode $j$ as
\begin{equation}
\hat{\Pi}_j^{\rm{off}}=(1-\nu)\ketbra{0}{0}_j
\end{equation}
and
\begin{equation}
\hat{\Pi}_j^{\rm{on}}=\hat{I}-\hat{\Pi}^{\rm{off}}_j,
\end{equation}
where $\nu$ is the dark-count probability. 
In the numerical simulation, $P^{\mathrm{suc}}$ is given by $P^{\mathrm{suc}}=P(\mathrm{D_{5H}}\cap \mathrm{D_{6V}})+P(\mathrm{D_{5V}}\cap \mathrm{\mathrm{D_{6H}}})$. 
In the experiment, the success probability of the BSM $P^{\mathrm{suc}}$ is equal to $P(\mathrm{D_{5V}}\cap \mathrm{\mathrm{D_{6H}}})$, since we only employ $\mathrm{D_{5V}}$ and $\mathrm{\mathrm{D_{6H}}}$. 
Hereafter, we consider the case where $P^{\mathrm{suc}}=P(\mathrm{D_{5V}}\cap \mathrm{\mathrm{D_{6H}}})$ for simplicity. 
The density operator of the heralded state~($\hat{\rho}^\mathrm{herald}_{H_1V_1H_4V_4}$) conditioned by the successful BSM is given by 
\begin{widetext}
\begin{align}
\hat{\rho}^{\mathrm{herald}}_{H_1V_1H_4V_4}
&=\frac{1}{P^{\mathrm{suc}}}\mathrm{Tr}_{\setminus H_1V_1H_4V_4}\left[\rho^{\gamma^{\mathrm{BSM}}_{\mathcal{U}}}(\hat{I}-(1-\nu)^3\ketbra{0}{0}^{\otimes3}_{V_3V_{3a}V_{3b}})(\hat{I}-(1-\nu)^3\ketbra{0}{0}^{\otimes3}_{H_2H_{2a}H_{2b}})\right]\\
&=\frac{1}{P^{\mathrm{suc}}}\sum_{i=0}^3(-1)^iP_i\mathrm{Tr}_{\setminus H_1V_1H_4V_4}[\hat{\rho}^{\gamma_i}]
\label{eq:swapchar}
\end{align}
\end{widetext}
where 
\begin{widetext}
\begin{eqnarray}
\hat{\rho}^{\gamma_0}&:=&\frac{1}{P_0}\hat{\rho}^{\gamma^{\mathrm{BSM}}_{\mathcal{U}}},\\
\hat{\rho}^{\gamma_1}&:=&\frac{1}{P_1}\mathrm{Tr}_{V_3V_{3a}V_{3b}}\left[\hat{\rho}^{\gamma^{\mathrm{BSM}}_{\mathcal{U}}}\ketbra{0}{0}^{\otimes3}_{V_3V_{3a}V_{3b}}\right],\\
\hat{\rho}^{\gamma_2}&:=&\frac{1}{P_2}\mathrm{Tr}_{V_3V_{3a}V_{3b}H_2H_{2a}H_{2b}}\left[\hat{\rho}^{\gamma^{\mathrm{BSM}}_{\mathcal{U}}}\ketbra{0}{0}^{\otimes6}_{V_3V_{3a}V_{3b}H_2H_{2a}H_{2b}}\right],\\
\hat{\rho}^{\gamma_3}&:=&\frac{1}{P_3}\mathrm{Tr}_{H_2H_{2a}H_{2b}}\left[\hat{\rho}^{\gamma^{\mathrm{BSM}}_{\mathcal{U}}}\ketbra{0}{0}^{\otimes3}_{H_2H_{2a}H_{2b}}\right].
\end{eqnarray}
\end{widetext}
Here, we define $\mathrm{Tr}_{\setminus H_1V_1H_4V_4}$ by partial trace over all remaining modes except for $H_1$, $V_1$, $H_4$ and $V_4$. 
The covariance matrices of $\hat{\rho}^{\gamma_1}$, $\hat{\rho}^{\gamma_1}$ and $\hat{\rho}^{\gamma_3}$ 
are given by the Schur complements~\cite{adesso2014continuous} of $\gamma^{\mathrm{BSM}}_{\mathcal{U}}$ as 
\begin{widetext}
\begin{align}
&\gamma_1:=\gamma^{\mathrm{BSM}}_{\{V_3V_{3a}V_{3b}\}\{V_3V_{3a}V_{3b}\}}-\gamma^{\mathrm{BSM}}_{\{V_3V_{3a}V_{3b}\}\{\mathcal{U}\setminus V_3V_{3a}V_{3b}\}}\left(\gamma^{\mathrm{BSM}}_{\{\mathcal{U}\setminus V_3V_{3a}V_{3b}\}\{\mathcal{U}\setminus V_3V_{3a}V_{3b}\}}\right. \nonumber\\
&\left.+I^{\oplus3}\right)^{-1}(\gamma^{\mathrm{BSM}}_{\{V_3V_{3a}V_{3b}\}\{\mathcal{U}\setminus V_3V_{3a}V_{3b}\}})^T,\\
&\gamma_2:=\gamma^{\mathrm{BSM}}_{\{V_3V_{3a}V_{3b}H_2H_{2a}H_{2b}\}\{V_3V_{3a}V_{3b}H_2H_{2a}H_{2b}\}}-\gamma^{\mathrm{BSM}}_{\{V_3V_{3a}V_{3b}H_2H_{2a}H_{2b}\}\{\mathcal{U}\setminus V_3V_{3a}V_{3b}H_2H_{2a}H_{2b}\}}\nonumber\\
&\times\left(\gamma^{\mathrm{BSM}}_{\{\mathcal{U}\setminus V_3V_{3a}V_{3b}H_2H_{2a}H_{2b}\}\{\mathcal{U}\setminus V_3V_{3a}V_{3b}H_2H_{2a}H_{2b}\}}\right. \nonumber\\
&\left.+I^{\oplus6}\right)^{-1}(\gamma^{\mathrm{BSM}}_{\{V_3V_{3a}V_{3b}H_2H_{2a}H_{2b}\}\{\mathcal{U}\setminus V_3V_{3a}V_{3b}H_2H_{2a}H_{2b}\}})^T,\\
&\gamma_3:=\gamma^{\mathrm{BSM}}_{\{H_2H_{2a}H_{2b}\}\{H_2H_{2a}H_{2b}\}}-\gamma^{\mathrm{BSM}}_{\{H_2H_{2a}H_{2b}\}\{\mathcal{U}\setminus H_2H_{2a}H_{2b}\}}\left(\gamma^{\mathrm{BSM}}_{\{\mathcal{U}\setminus H_2H_{2a}H_{2b}\}\{\mathcal{U}\setminus H_2H_{2a}H_{2b}\}}\right. \nonumber\\
&\left.+I^{\oplus3}\right)^{-1}(\gamma^{\mathrm{BSM}}_{\{H_2H_{2a}H_{2b}\}\{\mathcal{U}\setminus H_2H_{2a}H_{2b}\}})^T.
\end{align}
\end{widetext}
Here, $\gamma^{\mathrm{BSM}}_{\{i_1\ldots i_n\}\{j_1\ldots j_n\}}$ is the submatrix obtained by 
deleting rows corresponding to modes $i_1\ldots i_n$ and columns corresponding to modes 
$j_1\ldots j_n$ from $\gamma^{\mathrm{BSM}}_{\mathcal{U}}$. 
Then, the characteristic function of the heralded state is given by
\begin{widetext}
\begin{eqnarray}
\chi_{\hat{\rho}^{\mathrm{herald}}_{H_1V_1H_4V_4}}
=\frac{1}{P^{\mathrm{suc}}}\sum_{i=0}^3(-1)^iP_i\mathrm{exp}\left(-\frac{1}{4}\xi^T\gamma_{i,H_1V_1H_4V_4}\xi\right),   
\end{eqnarray}
\end{widetext}
where $\gamma_{i,H_1V_1H_4V_4}$ is the covariance matrix of $\mathrm{Tr}_{\setminus H_1V_1H_4V_4}[\hat{\rho}^{\gamma_i}]$. 
Before the detection, we perform the symplectic transformations of the (polarization-domain) beamsplitters 
followed by the detection losses on each of $\gamma_{i,H_1V_1H_4V_4}$ for $i\in\{0,1,2,3\}$ as 
\begin{widetext}
\begin{equation}
\gamma^{\mathrm{final}}_{i,H_1V_1H_4V_4}:=\mathcal{L}^{\eta_1}_{H_1}\mathcal{L}^{\eta_2}_{H_4}\mathcal{L}^{\eta_3}_{V_1}
\mathcal{L}^{\eta_4}_{V_4}\left[(S^{\theta_{A}}_{H_1V_1}\oplus S^{\theta_{B}}_{H_4V_4})^T\gamma_{i,H_1V_1H_4V_4}(S^{\theta_{A}}_{H_1V_1}\oplus S^{\theta_{B}}_{H_4V_4})\right], 
\end{equation}
\end{widetext}
where $\theta_A$ and $\theta_B$ are the measurement angles for Alice and Bob, respectively.  
Finally, we calculate the detection probabilities. 
For example, the probability of observing clicks in $\mathrm{\mathrm{D_1}}$ and $\mathrm{\mathrm{D_2}}$ and no-clicks 
in $\mathrm{\mathrm{D_3}}$ and $\mathrm{\mathrm{D_4}}$ 
under the condition of the above measurement angles~($=:P(\mathrm{c1,c2,nc3,nc4}|\theta_{A},\theta_{B})$) is given by 
\begin{widetext}
\begin{align}
&P(\mathrm{c1,c2,nc3,nc4}|\theta_{A},\theta_{B})=\frac{1}{P^{\mathrm{suc}}}\mathrm{Tr}\left[\hat{\Pi}^{\rm{on}}_{H_1}(\nu)\hat{\Pi}^{\rm{on}}_{H_4}(\nu)\hat{\Pi}^{\rm{off}}_{V_1}(\nu)\hat{\Pi}^{\rm{off}}_{V_4}(\nu)\sum_{i=0}^3(-1)^iP_i\hat{\rho}^{\gamma^{\mathrm{final}}_{i,H_1V_1H_4V_4}}\right]\\
&=\frac{1}{P^{\mathrm{suc}}}\mathrm{Tr}\left[\sum_{i=0}^3(-1)^iP_i\hat{\rho}^{\gamma^{\mathrm{final}}_{i,H_1V_1H_4V_4}}
(\hat{I}-(1-\nu)\ketbra{0}{0}_{H_1})\right.\nonumber\\
&\left.\times(\hat{I}-(1-\nu)\ketbra{0}{0}_{H_4})(1-\nu)\ketbra{0}{0}_{V_1}(1-\nu)\ketbra{0}{0}_{V_4}\right]\\
&=\frac{1}{P^{\mathrm{suc}}}\sum_{i=0}^3\left((-1)^iP_i\left(\frac{4(1-\nu)^2}{\sqrt{\mathrm{det}(\gamma_{i,V_1V_4}^{\mathrm{final}}+I)}}-\frac{8(1-\nu)^3}{\sqrt{\mathrm{det}(\gamma_{i,H_1V_1V_4}^{\mathrm{final}}+I)}}\right.\right.\nonumber\\
&\left.\left.-\frac{8(1-\nu)^3}{\sqrt{\mathrm{det}(\gamma_{i,V_1H_4V_4}^{\mathrm{final}}+I)}}+\frac{16(1-\nu)^4}{\sqrt{\mathrm{det}(\gamma_{i,H_1V_1H_4V_4}^{\mathrm{final}}+I)}}\right)\right), 
\end{align}
\end{widetext}
where $\gamma_{i,j_1\ldots j_n}$ is the submatrix 
obtained by extracting the rows and columns corresponding to modes $j_1\ldots j_n$ 
from $\gamma^{\mathrm{final}}_i$. For calculating $S$, we adopt the same rule as 
what described in the appendix of Ref.~\cite{PhysRevA.98.063842}. 

\subsection{The Characteristic Function Of The Fock States}
The characteristic function of the four-mode Fock state is represented by 
\begin{equation}
\chi_{\ketbra{klmn}{k'l'm'n'}}=\chi_{\ketbra{k}{k'}}\chi_{\ketbra{l}{l'}}\chi_{\ketbra{m}{m'}}\chi_{\ketbra{n}{n'}}. 
\end{equation}
Here, we only consider up to single-photon state for each mode i.e. 
$k,k',l,l',m,m',n,n'\in\{0,1\}$. 
The characteristic function of the single-mode state $\ketbra{n}{m}$ is given by 
the inner product with the displacement operator $\hat{D}(\alpha):=\mathrm{exp}(\alpha\hat{a}^\dagger-\alpha^*\hat{a})$ as~\cite{barnett2002methods} 
\begin{widetext}
\begin{eqnarray}
\chi_{\ketbra{n}{m}}&=&\mathrm{Tr}[\ketbra{n}{m}\hat{D}(\alpha)]\\
&=&\bra{m}\mathrm{exp}(\alpha\hat{a}^\dagger-\alpha^*\hat{a})\ket{n}\\
   &=& \left\{ \begin{array}{ll}
    \sqrt{\frac{n!}{m!}}\mathrm{exp}(-|\alpha|^2/2)(-\alpha)^{m-n}L^{(m-n)}_n(|\alpha|^2) & (m>n) \\
   \sqrt{\frac{m!}{n!}}\mathrm{exp}(-|\alpha|^2/2)(\alpha^*)^{n-m}L^{(n-m)}_m(|\alpha|^2) & (n>m),
  \end{array} \right.
\end{eqnarray}
\end{widetext}
where 
\begin{equation}
    L^{(k)}_{l}(x):=\sum_{i=0}^l(-1)^i \left(
    \begin{array}{c}
      l+k \\
      l-i 
    \end{array}
  \right)
  \frac{x^i}{i!}
\end{equation}
is the generalized Laguerre polynomial. 
We note that, in the single-mode case, the complex number $\alpha$ in the displacement operator and the complex numbers $\xi_1$ and $\xi_2$ in the Weyl operator are
connected by 
\begin{equation}
 \alpha=\frac{\xi_2-i\xi_1}{\sqrt{2}}.
\end{equation}

\subsection{Input State Characterization}
\begin{figure}[t]
 \begin{center}
 \includegraphics[width=\columnwidth]{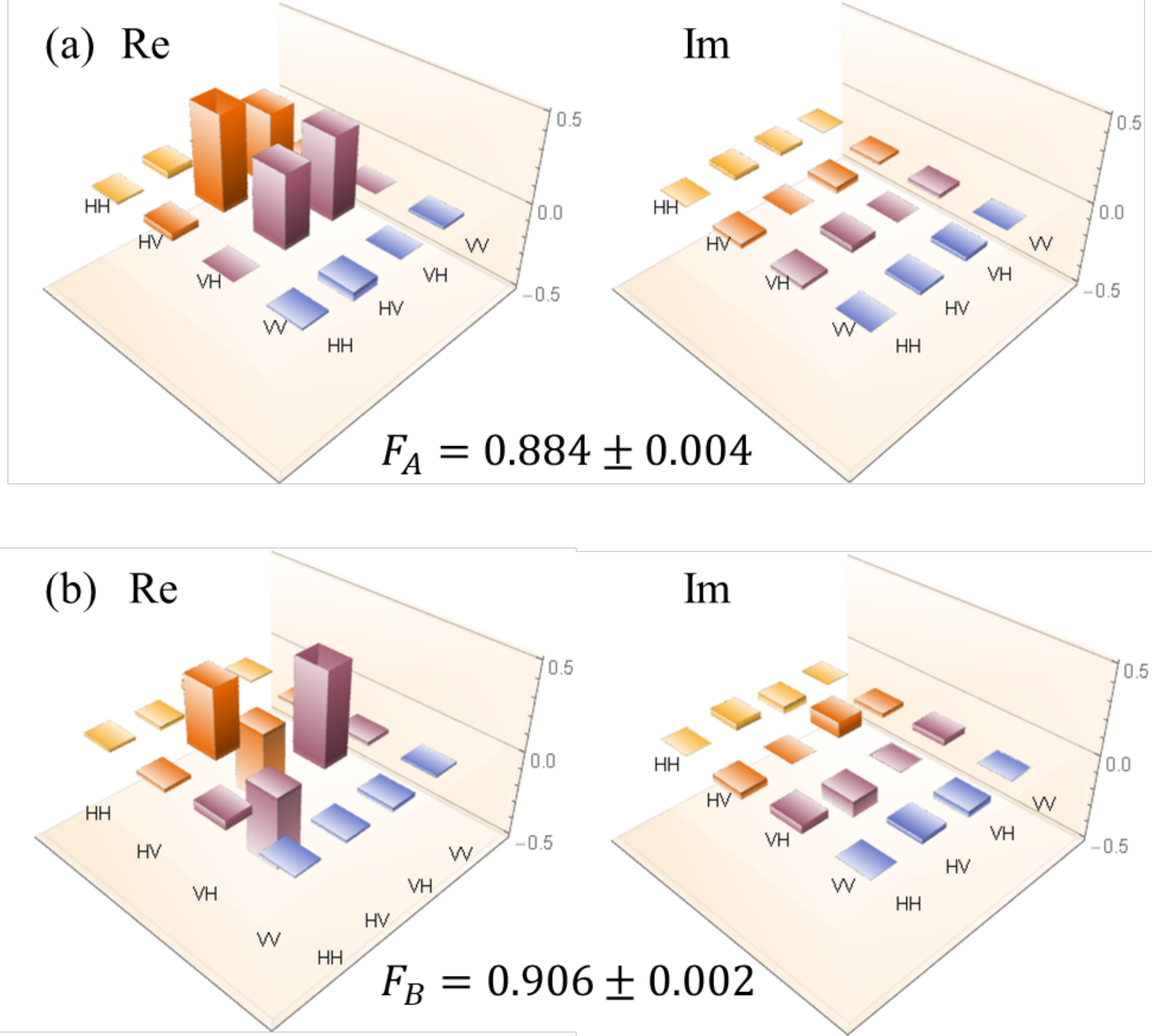}
  \caption{The real parts and imaginary parts of $\hat{\rho}_A$~(a) and $\hat{\rho}_B$~(b).} 
\label{fig:initial}
 \end{center}
\end{figure}

We characterize the input quantum states by performing the two-qubit quantum state tomography~\cite{PhysRevA.64.052312}. 
Changing the measurement angles,  
we collect the two-fold coincidence counts between $\mathrm{\mathrm{D_1}~(\mathrm{D_2})}$ and $\mathrm{\mathrm{\mathrm{D_{6H}}}}$ 
for characterizing the quantum state generated from source~A~(B), respectively, 
The two-qubit quantum states generated by the sources~A and B are 
reconstructed by performing the maximally likelihood estimation~\cite{PhysRevA.75.042108} using the 
probability distributions obtained by the experiment. 
The reconstructed two-qubit density operators generated from the sources~A~($\hat{\rho}_A$) and B~($\hat{\rho}_B$) are 
shown in Figs.~\ref{fig:initial}~(a) and (b), respectively.
The fidelity of $\hat{\rho}_A$ to $\ketbra{\Psi^+}{\Psi^+}$ is calculated to be $F_A:=\bra{\Psi^+}\hat{\rho}_A\ket{\Psi^+}=0.884\pm0.004$. 
Similarly, the fidelity of $\hat{\rho}_B$ to $\ketbra{\Psi^-}{\Psi^-}$ is calculated to be $F_B:=\bra{\Psi^-}\hat{\rho}_B\ket{\Psi^-}=0.906\pm0.002$. 
Theses results indicate that highly entangled states are prepared as initial states. 
The error bars are obtained by assuming a Poissonian distribution for the photon counts.

\subsection{Characterization of Indistinguishability}
\begin{figure}[t]
 \begin{center}
 \includegraphics[width=\columnwidth]{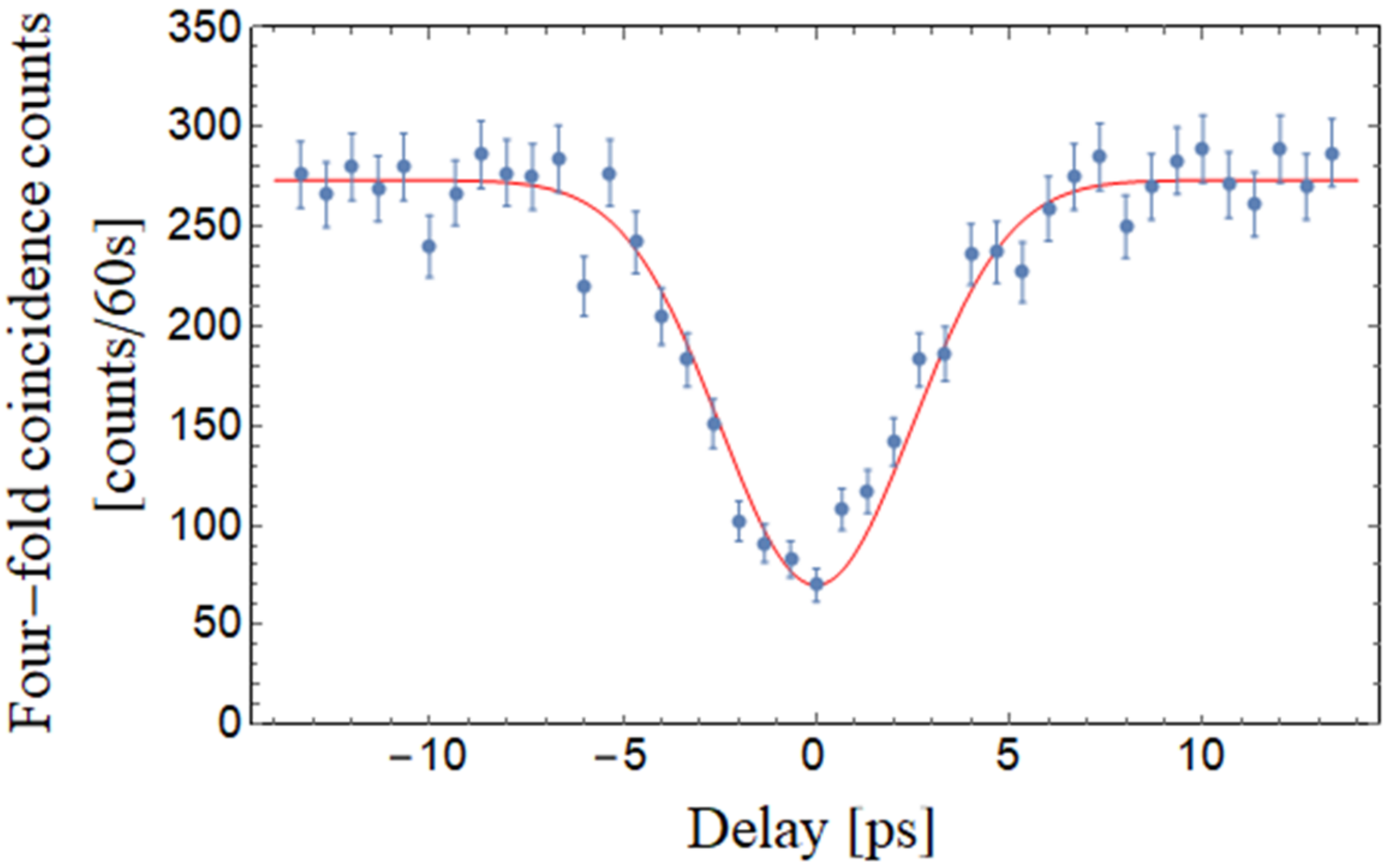}
  \caption{The observed HOM interference between the photon~3 and the photon~4. 
	The blue dots are the four-fold coincidence counts in 60 seconds. The error bars are 
	calculated by assuming a Poissonian distribution. The red solid curve is obtained by Gaussian fitting. }	
\label{fig:HOM}
 \end{center}
\end{figure}

In order to evaluate the indistinguishability between the photon~3 and the photon~4 which interfere at the HBS, 
we perform the HOM experiment~\cite{PhysRevLett.59.2044,Tsujimoto:15,PhysRevA.87.063801}. 
We detect the photons~1 and 2 with $V$-polarization, 
and observe the HOM interference between the $H$-polarized photons in modes 3 and 4. 
We measure the four-fold coincidence counts among $\mathrm{\mathrm{D_1}, \mathrm{D_2}, \mathrm{D_{5V}}}$, and $\mathrm{\mathrm{\mathrm{D_{6H}}}}$ 
with changing the relative delay by means of a motion stage. 
The result is shown in Fig.~\ref{fig:HOM}. 
We clearly see the HOM dip around the zero-delay point. 
The visibility is calculated to be $V_{\mathrm{HOM}}=0.74\pm0.03$. 
The degradation of the visibility is mainly caused by 
(i)~The mode matching~$T_\mathrm{mode}$ between the photons~3 and 4, 
and (ii)~Multiple pair generation at the sources. 
To see the degree of the contribution of $T_\mathrm{mode}$, 
we perform the theoretical calculation considering the experimental 
imperfections. 
When we set $T_\mathrm{mode}=1$, the visibility is estimated to be $V^{\mathrm{th}}_{\mathrm{HOM}}=0.91$,  
which indicates that the remaining degradation is caused by the mode mismatch. 
$V^{\mathrm{th}}_{\mathrm{HOM}}=0.74$ is obtained for $T_\mathrm{mode}=0.9$. 
We adopt this value in the numerical simulations. 

 \begin{figure}[t]
 \begin{center}
 \includegraphics[width=\columnwidth]{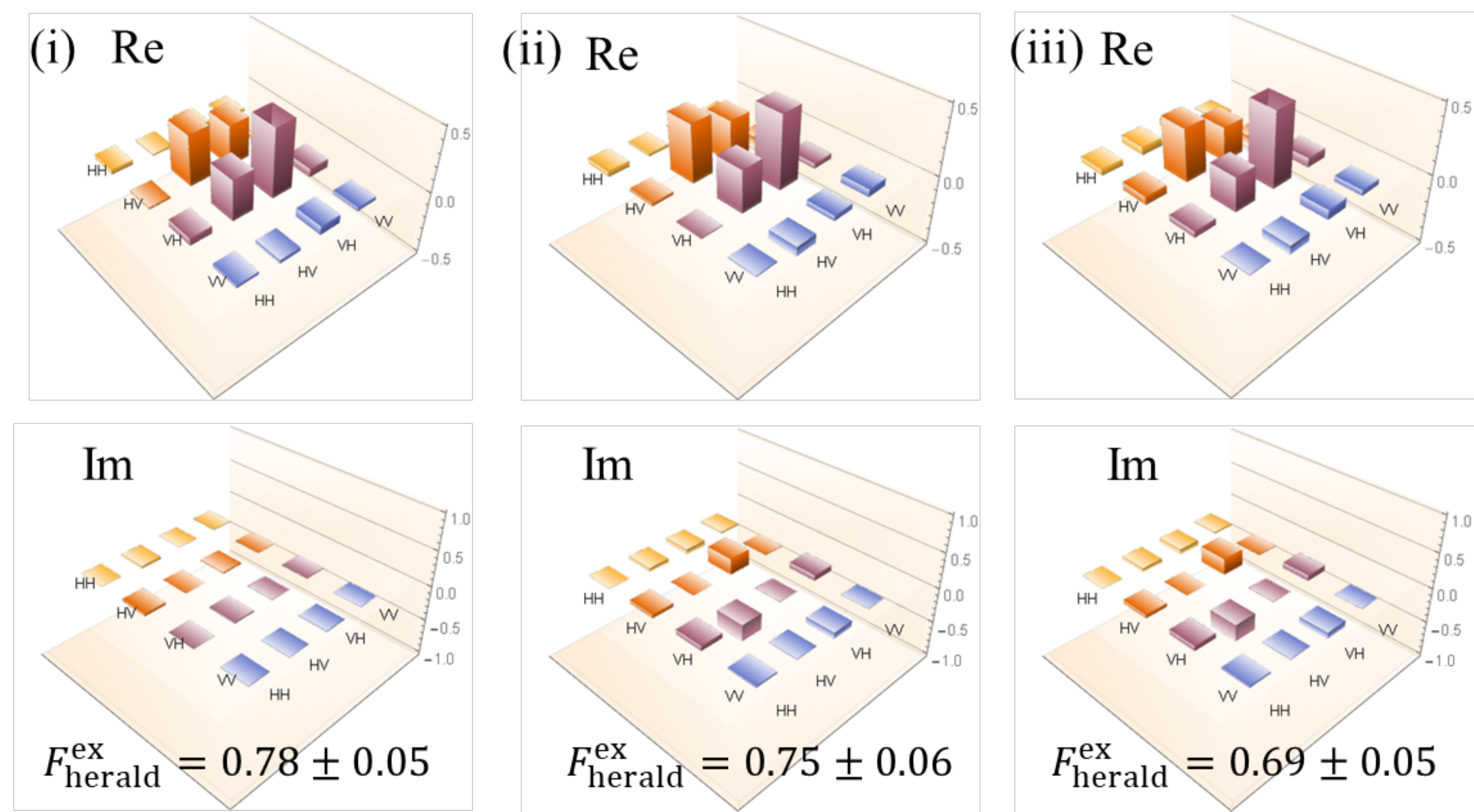}
  \caption{The reconstructed density operators of the two-qubit component of the heralded state for three different distances: 
  (i)~0~km, (ii)~24~km and (iii)~50~km.}	\label{fig:swap2}
 \end{center}
\end{figure}

\subsection{Characterization of The Heralded State}
We show the two-qubit density operators of the heralded states 
reconstructed by the experimentally-obtained probability distributions in Fig.~\ref{fig:swap2}. 
(i), (ii) and (iii) correspond to the two-qubit density operators of the 
heralded states when 
the equivalent-fiber-lengths are 0~km, 24~km and 50~km, respectively. 
The fidelities to $\ket{\Psi^+}$ 
are calculated to be (i)~$F^{\mathrm{ex}}_{\mathrm{herald}}=0.78\pm0.05$, (ii)~$F^{\mathrm{ex}}_{\mathrm{herald}}=0.75\pm0.06$, and (iii)~$F^{\mathrm{ex}}_{\mathrm{herald}}=0.69\pm0.05$, respectively. 
In theory, the fidelities are estimated to be $F^{\mathrm{th}}_{\mathrm{herald}}=0.81$ regardless of the distance. 
We guess the reason why $F^{\mathrm{ex}}_{\mathrm{herald}}$ is lower than $F^{\mathrm{th}}_\mathrm{herald}$ is 
that additional spatial mode-mismatch is caused by inserting ND filters. 

\section*{Funding.}
Core Research for Evolutional Science and
Technology, Japan Science and Technology Agency~(CREST) 
(JPMJCR1772); 
Ministry of Education, Culture, Sports, Science, and
Technology~(MEXT); Japan Society for the Promotion of
Science~(JSPS) (JP18K13487, JP17K14130, JP17K05091); 
The US National Science Foundation.

\section*{Acknowledgment.}
We thank Sushovit~Adhikari, Kaushik~P.~Seshadreesan and Rikizo Ikuta for helpful discussions.

\end{document}